\documentclass{LMCS}

\usepackage{textcomp}
\usepackage{pifont}
\usepackage{latexsym}
\usepackage{amsfonts}
\usepackage{amssymb}
\usepackage{amsmath}
\usepackage{stmaryrd}
\usepackage[T1]{fontenc}
\usepackage[english]{babel}
\usepackage{hyperref}
\usepackage{enumerate}
\usepackage{url}
\usepackage{array}
\usepackage[numbers]{natbib}
\usepackage{tikz}
\usetikzlibrary{arrows,decorations,decorations.pathmorphing,positioning,shapes,backgrounds,fit}
\usepackage{arxiv}

\theoremstyle{plain}
\newtheorem{claim}[thm]{Claim}
\newtheorem*{theorem*}{Theorem}
\newenvironment{thmcont}[1]{\em\trivlist\item[\hskip\labelsep
   {\bf Theorem~\ref{#1} (revisited).}]}{\endtrivlist}
\theoremstyle{definition}
\newtheorem*{notation}{Notation}

\def\doi{7 (3:01) 2011}
\lmcsheading%
{\doi}
{1--38}
{}
{}
{Oct.~11, 2010}
{Jul.~12, 2011}
{}   

\begin{document}

\title{The Derivational Complexity Induced by the Dependency Pair Method}

\author[G.~Moser]{Georg Moser\rsuper a} 
\address{Institute of Computer Science, University of Innsbruck, Austria} 
\email{georg.moser@uibk.ac.at, andreas.schnabl@uibk.ac.at}  
\thanks{{\lsuper{a,b}}Supported by FWF (Austrian Science Fund) project P20133-N15} 

\author[A.~Schnabl]{Andreas Schnabl\rsuper b} 
\address{\vskip-6 pt} 
\thanks{{\lsuper b}Supported by a grant of the University of Innsbruck} 

\titlecomment{A preliminary version of this article appeared as \cite{MS09}}

\begin{abstract}
We study the derivational complexity induced by the dependency pair
method, enhanced with standard refinements. We obtain upper bounds
on the derivational complexity induced by the dependency pair method 
in terms of the derivational complexity of the base techniques employed.
In particular we show that the derivational complexity
induced by the dependency pair method based on some direct technique, possibly
refined by argument filtering, the usable rules criterion, or dependency
graphs, is primitive recursive in the derivational complexity induced by the
direct method.
This implies that the derivational complexity induced
by a standard application of the dependency pair method based on traditional
termination orders like KBO, LPO, and MPO is exactly the same as if those orders
were applied as the only termination technique.
\end{abstract}

\maketitle

\section{Introduction}
\label{sec:introduction}

Several notions to assess the complexity of a 
terminating term rewrite system (TRS) have been proposed in the literature, 
compare~\cite{CKS:1989,CL:1992,HM08a,HL89}.
The conceptually simplest one was suggested by Hofbauer and Lautemann in~\cite{HL89}: 
the complexity of a given TRS is measured as the maximal length of derivation sequences. 
More precisely, the \emph{derivational complexity function} with respect to a terminating TRS $\RS$
relates the maximal derivation height to the size of the initial term.
We adopt this notion as our central definition of the complexity of a TRS. 

For termination proofs by direct methods a considerable 
number of results establish
essentially optimal upper bounds on the growth rate of the derivational complexity function.
See for example~\cite{H92,HL89,L01,MS08,MW03,MSW08,NZM:2010,W:2010,W95} for 
results in this direction.
However, for transformation techniques like semantic labelling~\cite{Z95}
or the dependency pair method~\cite{AG00} the situation changes.
For semantic labelling, it is a trivial observation that the derivational complexity of the
original TRS is bounded from above by the derivational complexity of the labelled system.
However, if the domain of the used (quasi-)models is infinite, the labelled TRS is generally
infinite, as well. Estimating the derivational complexity of such systems is harder than for
finite systems: for some termination proof methods, such as the multiset path order (MPO for
short) or the lexicographic path order (LPO for short)
the complexity results only hold for finite TRSs \cite{H92,W95}. For the
Knuth-Bendix order (KBO for short) the situation is better. If some weak conditions
are in place, then the bound on the derivational complexity with respect to
finite TRSs extends to infinite TRSs~\cite{M06}.
With respect to the dependency pair method, 
in~\cite{AM09,HM08a,HM08b,MP08,MP09,NEG11,ZK10}  the bounds on
derivation heights induced by the dependency pair method or
its framework are investigated. However only variations on the
original definition of the dependency pair method were analysed.

In this paper we give a derivational complexity analysis
of the dependency pair method. It should be emphasised 
that the notion of dependency pair method studied here amounts to the
original technique as introduced by Arts and Giesl~\cite{AG00} 
(see also Hirokawa and Middeldorp~\cite{HM05}). 
As the dependency pair method is a transformation technique, we
can only give a parametrised analysis. We call those techniques
that are applied on the transformed system: \emph{base} techniques.
Let us exemplify this notation on the next example.

\begin{exa}
\label{ex:intro}
Consider the TRS $\RSa$ given below:
\begin{alignat*}{4}
&& \mi(x) \mcirc (y \mcirc z) &\rew \mf(x,\mi(x)) \mcirc (\mi(\mi(y)) \mcirc z) & \hspace{10ex}
&& \mi(x) &\rew x\\
&& \mi(x) \mcirc (y \mcirc (z \mcirc w)) &\rew \mf(x,\mi(x)) \mcirc (z \mcirc (y \mcirc w)) & \hspace{10ex}
&& \mf(x,y) &\rew x \tpkt
\end{alignat*}
$\RSa$ is a variation of a TRS encoding the Ackermann function,
introduced by Hofbauer~\cite[Proposition 5.9]{HL89} (also compare~\cite{H92b}). 
Note that $\RSa$ is not simply terminating
and the derivational complexity of $\RSa$ grows as fast as
the Ackermann function.
However, termination can be (automatically) 
shown by the dependency pair method in 
conjunction with argument filtering and~KBO
(we give all necessary definitions in Sections~\ref{sec:preliminaries}
and~\ref{sec:DPMethod}).
\end{exa}

In Example~\ref{ex:intro} we cannot apply KBO directly (the TRS $\RSa$ 
is not simply terminating), but we apply KBO as base technique. In order
to measure the strength of the dependency pair method itself, we 
express the induced complexity relative to the (maximal) complexities
of the base techniques. With respect to Example~\ref{ex:intro} it
is not difficult to see that the derivational complexity of
$\RSa$ belongs to $\Ack(\Theta(n),0)$, where $n$ is the size of the 
start term. Essentially this follows from~\cite[Proposition 5.9]{H92b}, 
due to the closeness of $\RSa$ to Hofbauer's original example. 
As this is also the complexity induced by KBO~\cite{L01} 
it may appear that the dependency pair method does not
add any power. Our results provide a clear picture of the true connection.
With respect to upper bounds on the derivational
complexity, we establish the following (technical) results:
\begin{enumerate}[(1)]
\item
\label{intro:dp}
For the basic dependency pair method (potentially using argument
filterings) the induced derivational complexity is bounded triple
exponentially in the derivational complexity of the base technique 
used.
If we restrict to string rewrite systems, then the induced 
derivational complexity is exponential in the derivational complexity of the base technique.
\item
\label{intro:ur}
If we consider the basic dependency pair method using
the usable rules refinement, then the induced derivational complexity
is primitive recursive in the derivational complexity of the base technique.
\item
\label{intro:dg}
Finally, if we consider the dependency pair method in conjunction
with dependency graphs, then the induced derivational complexity is primitive
recursive in the (maximal) derivational complexity of the base techniques employed.
\end{enumerate}
Complementing these results, we present results on lower bounds.
For the basic dependency pair method, we present an example which shows
that at least two of the three exponentials in its upper bound can actually
be reached. If we restrict to string rewriting, this bound reduces to a
single exponential. Hence the corresponding upper bound mentioned in result~(\ref{intro:dp})
is optimal. 
For the usable rules refinement, we show that the growth rate
of the derivational complexity function may be nonelementary. Furthermore
we show that the bound for dependency graphs given by result~(\ref{intro:dg})
is essentially optimal.

To exemplify these results, we momentarily focus on polynomial 
interpretations as base technique.
It is well-known that polynomial interpretations induce a double exponential
bound on the derivational complexity~\cite{HL89}. Let $\RS$ be a
TRS and suppose termination of $\RS$ has been established by applying the basic
dependency pair method, where polynomial interpretations are used to define the
employed reduction pair. According to result~(\ref{intro:dp})
the derivational complexity function with respect to $\RS$ is bounded
by $2_5(\bO(n))$, i.e, by a tower of $2$s of height $5$ in $n$.
On the other hand, if in addition the usable rules criterion or 
dependency graphs are used, 
then results~(\ref{intro:ur}) and~(\ref{intro:dg}) yield
that the derivational complexity is (at most) primitive recursive.

Thus seemingly easy refinements of
the dependency pair method like dependency graphs may lead to noteworthy
speed-ups of the growth rates of the derivational complexity function. 
On the other hand if strong techniques (with respect to the complexity induced) are employed in conjunction with the dependency pair method, then 
the derivational complexity of the analysed TRS may only depend on 
the complexity induced by the base technique. 

Re-consider the TRS $\RSa$ given in Example~\ref{ex:intro}. 
There are nine dependency pairs.
\begin{alignat*}{4}
&& \mi(x) \mcirc^\sharp (y \mcirc z) &\rew \mf(x,\mi(x)) \mcirc^\sharp (\mi(\mi(y)) \mcirc z) & \hspace{2ex}
&& \mi(x) \mcirc^\sharp (y \mcirc (z \mcirc w)) &\rew \mf(x,\mi(x)) \mcirc^\sharp (z \mcirc (y \mcirc w))
\\
&& \mi(x) \mcirc^\sharp (y \mcirc z) &\rew \mf^\sharp(x,\mi(x)) &
&& \mi(x) \mcirc^\sharp (y \mcirc (z \mcirc w)) &\rew \mf^\sharp(x,\mi(x))
\\
&& \mi(x) \mcirc^\sharp (y \mcirc z) &\rew \mi(\mi(y)) \mcirc^\sharp z &
&& \mi(x) \mcirc^\sharp (y \mcirc (z \mcirc w)) &\rew z \mcirc^\sharp (y \mcirc w)
\\
&& \mi(x) \mcirc^\sharp (y \mcirc z) &\rew \mi^\sharp(\mi(y)) &
&& \mi(x) \mcirc^\sharp (y \mcirc (z \mcirc w)) &\rew y \mcirc^\sharp w
\\
&& \mi(x) \mcirc^\sharp (y \mcirc z) &\rew \mi^\sharp(y)
\tpkt
\end{alignat*}
To show termination of $\RSa$ one may use
the argument filtering $\pi$: $\pi(\mf)=\pi(\mf^\sharp)=\pi(\mi^\sharp)=1$,
$\pi(\mi)=[1]$, $\pi(\mcirc)=\pi(\mcirc^\sharp)=[1,2]$
and the reduction pair $(\geqkbopi,\gkbopi)$, where $(\geqkbopi,\gkbopi)$
is induced by the (admissible) weight function $\m{w}$ with $w_0=1$, 
$\m{w}(\mcirc)=\m{w}(\mcirc^\sharp)=1$, and $\m{w}(\mi)=0$. Furthermore 
the precedence $\succ$ fulfils: $\mi\succ\;\mcirc,\mcirc^\sharp$.
Due to result~(\ref{intro:dp}) and~\cite{L01} the derivational complexity of $\RSa$ belongs to $\Ack(\Theta(n),0)$.

In contrast to the case for polynomial interpretations, the 
complexity induced by the base technique belongs to a class of
functions closed under primitive recursion. Hence it is
already so huge, that the inherent complexity of 
the dependency pair method becomes negligible.

Note the challenges of our investigation:
In order to estimate the derivational complexity of a rewrite system
we only consider the (maximal) derivation complexities
induced by the base techniques employed. This entails that we exploit
the upper bound on the maximal number of dependency pair steps to bound the length of derivations.

\medskip
Some of the results in this paper appeared in an earlier conference paper~\cite{MS09}. 
Apart from correcting some shortcomings of the
conference paper the journal version extends~\cite{MS09} by providing
a full analysis of the dependency graph refinement (see Section~\ref{sec:dg}).
Furthermore the treatment of the usable rules criterion is new (see Section~\ref{sec:ur}), as well as the improvement of the lower- and upper-bound in the context of
string rewriting (see Section~\ref{sec:srscomplexity}).

The technically most involved result is the proof of the triple exponential upper bound for
the basic dependency pair method. Our proof rests on the observation that it suffices
to bound the maximal depth of a term during a given derivation. 
We show that the depth of any term occurring in a derivation is bounded exponentially in
the maximal number of dependency pair steps. Based on this result the triple exponential
upper bound follows by standard observations.
Due to this ground work the analysis of the usable rules refinement is relatively
straightforward.
On the other hand, for the analysis of the dependency graph refinement we employ a
different, but conceptually simpler technique. Essentially, it suffices to
embed the analysed TRS in a generic simulating TRS whose derivational complexity
can be analysed directly.

The rest of this paper is organised as follows. In Sections~\ref{sec:preliminaries}
and~\ref{sec:DPMethod} we present basic notions and starting points of the paper.
Sections~\ref{sec:ancestors} and~\ref{sec:dpcomplexity} establish
result~(\ref{intro:dp}). The mentioned improvement for string rewriting
is given in Section~\ref{sec:srscomplexity}. In Section~\ref{sec:ur}, we extend our considerations to
usable rules and thus show result~(\ref{intro:ur}). 
In Section~\ref{sec:dg}, we consider dependency graphs and show
result~(\ref{intro:dg}). Finally we conclude in Section~\ref{sec:conclusion}.
To ease the presentation some technical results have been moved to the appendix.

\section{Preliminaries}
\label{sec:preliminaries}

We assume familiarity with the basics of term rewriting, see~\cite{BN98,Terese}.
Below we recall central definitions and notions of rewriting which are relevant
to this paper.

Let $\VS$ denote a countably infinite set of variables and $\FS$ a signature of
function symbols with fixed arities. The set of terms over $\FS$ and $\VS$ is
denoted as $\TERMS$. The set of ground terms over $\FS$
is denoted as $\TA(\FS)$.
The (proper) subterm relation is denoted as $\subterm$ ($\prsubterm$);
we write $\superterm$ ($\prsuperterm$) for the reversed (proper) subterm
relation.
The \emph{root symbol} (denoted as $\rt(t)$) of a term $t$ is either $t$ itself, if
$t \in \VS$, or the symbol $f$, if $t = f(\seq{t})$.
We denote the set of variables occurring in a term $t$ as $\Var(t)$, and the
set of function symbols occurring in $t$ as $\Fun(t)$.
A \emph{position} is a finite sequence of positive integers. The root position is
the empty sequence denoted by $\epsilon$, and $pq$ denotes the concatenation of positions
$p$ and $q$. The set of positions of a term $t$ is denoted as $\Pos(t)$.
We write $p\leqslant q$ ($p<q$) to denote that $p$ is a (proper) prefix of $q$,
and $p\parallel q$ if neither $p\leqslant q$ nor $q \leqslant p$.
The subterm of $t$ at position $p$ is denoted as $\atpos{t}{p}$.
We write $\FPos(t)$ ($\VPos(t)$) for the set of positions $p$ such that $\FS$ ($\VS$)
contains $\rt(\atpos{t}{p})$.
To simplify the exposition, we often confuse terms and their tree
representations: a \emph{branch} of a term $t$ is a maximal set of positions
$B$ in $t$ such that for all pairs of positions $q,q'\in B$, we have
$q\leqslant q'$ or $q'\leqslant q$.
The \emph{size} (denoted as $\size{t}$) of a term $t$ is the
number of variables and function symbols occurring in $t$. The
\emph{depth} (denoted as $\depth{t}$) of a term $t$ is $0$ if $t$
is a variable or a constant, and defined as follows if $t=f(t_1,\dots,t_n)$:
$\depth{t} \defsym 1 + \max \{ \depth{t_i} \colon 1 \leqslant i \leqslant n\}$.
A \emph{substitution} is a mapping $\sigma:\VS\rightarrow\TERMS$.
The result of applying a substitution $\sigma$ to a term $t$ is denoted
as $t\sigma$.
We introduce a fresh constant $\ctx$ (the \emph{hole}) and define a \emph{context}
$C$ as a term (over $\FS \cup \{\ctx\}$ and $\VS$) containing exactly
one $\ctx$. For a term $t$
and a context $C$, $C[t]$ denotes the replacement of $\ctx$ by $t$.

A \emph{term rewrite system} (\emph{TRS} for short) $\RS$ over $\TERMS$ is a \emph{finite} set of rewrite rules $l\rew r$
with $l,r\in\TERMS$, $l\notin\VS$, and $\Var(r)\subseteq\Var(l)$. 
Given a TRS $\RS$ and two terms $s,t$, we say that $s$ \emph{rewrites} to
$t$ (denoted as $s\rsrew{\RS}t$) if there exist
a context $C$, a substitution $\sigma$ and a rewrite rule $l\rew r$ in $\RS$ such that
$s=C[l\sigma]$ and $t=C[r\sigma]$. If no confusion can arise, we write $s\rew t$, instead.
We write $\rstrew{\RS}$ for the transitive closure of this relation.
The reflexive closure is $\rsgrew{\RS}$.
The reflexive and transitive closure is denoted as $\rssrew{\RS}$. We write
$\rew_{\RS}^n$ to express $n$-fold composition of $\rsrew{\RS}$. If we wish to
indicate the redex position $p$ and the applied rewrite rule $l \rew r$ in a
reduction from $s$ to $t$, we write $s \completerew{p}{l \rew r} t$. A TRS $\RS$ is
\emph{terminating} if there exists no infinite chain of terms $t_0,t_1,\ldots$ such
that $t_i\rsrew{\RS}t_{i+1}$ for each~$i\in\N$.

A function symbol $f$ is \emph{defined} if $f=\rt(l)$ for some rewrite rule $l\rew r$
in the considered TRS $\RS$, otherwise it is called a \emph{constructor}.
The set of defined function symbols of $\RS$ is denoted as $\DS_\RS$, while the constructor
symbols are collected in $\CS_\RS$ (we only write $\DS$ and $\CS$, respectively, if no
confusion can arise).
We write $\DPos(t)$ ($\CPos(t)$) for the set of positions $p$ such that $\DS$ ($\CS$) 
contains $\rt(\atpos{t}{p})$.
We recall the notion of \emph{relative rewriting}, c.f.~\cite{G90,Terese}.
Let $\RR$ and $\SS$ be TRSs.
We write $\rsrew{\RR / \SS}$ for $\rssrew{\SS} \cdot \rsrew{\RS} \cdot \rssrew{\SS}$ and
we call $\rsrew{\RR / \SS}$ the \emph{relative rewrite relation} of $\RR$ modulo $\SS$.
Clearly, we have that ${\rsrew{\RR / \SS}} = {\rsrew{\RS}}$, if $\SS = \varnothing$.
We write $\NF(\RS)$, $\NF(\RS/\SS)$
to denote the set of normal forms of $\rsrew{\RS}$, $\rsrew{\RS/{\SS}}$ respectively.

The \emph{derivation height} of a term $s$ with respect to a
finitely branching, well-founded binary relation $\rew$ on terms is defined as
$\dheight(s,\rew) \defsym \max\{ n \mid \exists t \; s \rew^n t \}$.
The \emph{derivational complexity function}
of $\RS$ is defined as:
\begin{equation*}
  \Dc{\RS}(n) \defsym \max\{\dheight(t,\rsrew{\RS}) \mid \size{t} \leqslant n\} \tpkt
\end{equation*}

In analogy to the mapping $\dheight$ we define functions tracing the depth or size of reducts.
The \emph{potential depth} of a term $s$ with respect to $\rew$
is defined as follows: $\pdp(s,\rew) \defsym \max\{ \depth{t} \mid s \rss t\}$; the
\emph{potential size} is defined by $\psz(s,\rew) \defsym \max\{ \size{t} \mid s \rss t\}$.
If termination of $\RS$ by some termination proof technique implies an upper
bound on $\Dc{\RS}$, we call that bound the derivational complexity \emph{induced}
by that technique, or simply the derivational complexity of that technique.

An \emph{$\FS$-algebra} $\AA$ for a signature $\FS$ consists
of a \emph{carrier} $A$ and, for every function symbol $f\in\FS$, an
\emph{interpretation function} $f_{\AA}: A^n\to A$, where $n$ is the
arity of $f$.
Given an \emph{assignment} $\alpha:\VS\to A$, we denote
the evaluation of a term $t$ in $\AA$ by
$\eval{\alpha}{t}$.
A \emph{monotone $\FS$-algebra} is a pair $(\A,\succ)$
where $\A$ is an $\FS$-algebra and $\succ$ is a proper
order such that for every function symbol $f\in\FS$, $f_\AA$
is strictly monotone in all coordinates with respect to $\succ$.
A \emph{weakly monotone $\FS$-algebra} $(\A,\succcurlyeq)$ 
is defined similarly, but for every function symbol $f\in\FS$, 
it suffices that $f_\AA$ is monotone in all coordinates 
(with respect to the preorder $\succcurlyeq$). A monotone $\FS$-algebra
$(\A,\succ)$ is called \emph{well-founded} if $\succ$ is
well-founded. Similarly, a weakly monotone $\FS$-algebra
$(\A,\succcurlyeq)$ is well-founded, if the proper order $\succ$
induced by $\succcurlyeq$ is well-founded.
Any well-founded monotone $\FS$-algebra $(\A,\succ)$ induces
a reduction order $\succ_\A$ on terms: define $s \succ_\A t$ if and only if
$\eval{\alpha}{s} \succ \eval{\alpha}{t}$ for all assignments $\alpha$.
We say $(\A,\succ)$ is \emph{compatible}
with a TRS $\RS$ if ${\RS} \subseteq {\succ_\A}$.
Similarly, given a weakly monotone algebra $(\A,\succcurlyeq)$, we define
$s \succcurlyeq_\A t$ if and only if $\eval{\alpha}{s} \succcurlyeq \eval{\alpha}{t}$,
and $s \succ_\A t$ if and only if $\eval{\alpha}{s} \succ \eval{\alpha}{t}$ for
all assignments $\alpha$.
A \emph{polynomial interpretation}  is an
interpretation into a well-founded monotone (weakly monotone)
algebra $(\A,>)$ ($(\A,\geqslant)$) such that 
$A\subseteq\N$, $>$ ($\geqslant$) 
is the standard strict order (preorder) on the natural numbers, and $f_\AA$ is
a polynomial for every function symbol $f$~\cite{L79}.

We briefly recall the definition of the class of
\emph{primitive recursive functions}. The following
number-theoretic functions are \emph{initial}:
(i) the constant zero functions of all arities: $\mz_n(x_1,\ldots,x_n)=0$,
(ii) the successor function $\ms(x)=x+1$, and 
(iii) the projection functions $\pi_i^n(x_1,\ldots,x_n)=x_i$.
A class $\CC$ of number-theoretic functions is 
\emph{closed under composition} if for all $m$-ary $g\in\CC$ and 
$n$-ary $h_1,\ldots,h_m\in\CC$, the function
\begin{equation*}
  f(x_1,\ldots,x_n)=g(h_1(x_1,\ldots,x_n),\ldots,h_m(x_1,\ldots,x_n)) \tkom
\end{equation*}
is contained in $\CC$, as well.
It is \emph{closed under primitive recursion} if for all 
$n$-ary $g\in\CC$ and $n+2$-ary $h\in\CC$ the following function 
$f$ is contained in $\CC$, as well:
\begin{align*}
f(0,x_1,\ldots,x_n)&=g(x_1,\ldots,x_n)\\
f(y+1,x_1,\ldots,x_n)&=h(f(y,x_1,\ldots,x_n),y,x_1,\ldots,x_n)
\tpkt
\end{align*}

The class of \emph{primitive recursive functions} is the smallest set of
number-theoretic functions which contains all initial functions and is closed
under composition and primitive recursion. The definition schemata for primitive
recursive functions can be translated to rewrite rules in the obvious way, 
see for example~\cite[Definition~2.6]{CW97}.

The $i^{\text{th}}$ iterate of a unary function $f$ is
denoted as $f^i$, a similar notation is used for the $i^{\text{th}}$ iterate
of a function symbol. Finally, we define the function $2_n$ as follows:
\begin{equation*}
  2_0(m) \defsym m \qquad 2_{n+1}(m) \defsym 2^{2_n(m)} \tpkt
\end{equation*}

\section{Dependency Pair Method}
\label{sec:DPMethod}

We recall the central notions of the dependency pair method~\cite{AG00,HM05}.
Let $t$ be a term. We set $t^\sharp \defsym t$ if $t \in \VV$, and 
$t^\sharp \defsym f^\sharp(t_1,\dots,t_n)$ if $t = f(\seq{t})$.
Here $f^\sharp$ is a new $n$-ary function symbol called 
\emph{dependency pair symbol}. For a signature
$\FS$, we define $\FS^\sharp \defsym \FS \cup \{f^\sharp \mid f\in \FS\}$.
The set $\DP(\RS)$ of  \emph{dependency pairs} of a TRS $\RS$
is defined as  $\{ l^\sharp \to u^{\sharp} \mid l \to r \in \RS,
u \subterm r, \rt(u) \in \DS, u \nprsubterm l \}$. We recall
the following characterisation of termination of a TRS.
\begin{prop} \label{p:DPbasic}
A TRS $\RS$ is terminating if and only if there exists no infinite derivation of the
form $t_1^\sharp \rssrew{\RS} t_2^\sharp \rsrew{\DP(\RS)} t_3^\sharp \rssrew{\RS} \ldots$
such that for all $i>0$, $t_i^\sharp$ is terminating with respect to $\RS$.
\qed
\end{prop}
An \emph{argument filtering} (for a signature $\FS$) is a
mapping $\pi$ that assigns to every $n$-ary function symbol $f \in \FS$
an argument position $i \in \{ 1, \dots, n \}$ or a (possibly empty)
list $[ \seq[m]{i} ]$ of argument positions with
$1 \leqslant i_1 < \cdots < i_m \leqslant n$.
The signature $\FSpi$ consists of all function symbols $f$ such that
$\pi(f)$ is some list $[ \seq[m]{i} ]$, where in $\FSpi$ the arity of
$f$ is $m$. Every argument filtering $\pi$ induces a mapping from
$\TERMS$ to $\TERMSpi$, also denoted by $\pi$:
\begin{equation*}
\pi(t) \defsym \begin{cases}
t & \text{if $t$ is a variable} \\
\pi(t_i) & \text{if $t = f(\seq{t})$ and $\pi(f) = i$} \\
f(\pi(t_{i_1}),\dots,\pi(t_{i_m})) &
\text{if $t = f(\seq{t})$ and $\pi(f) = [ \seq[m]{i} ]$}
\tpkt
\end{cases}
\end{equation*}
An argument filtering $\pi$ is extended in the usual way
to a TRS $\RS$. Let $R$ be a binary relation, then we
write ${\pi(\RS)} \subseteq {R}$ to indicate that for all
${l \to r} \in {\RS}$, ${\pi(l)} \mathrel{R} {\pi(r)}$ holds.

A \emph{reduction pair} $(\succcurlyeq, \succ)$ consists of a
preorder $\succcurlyeq$ which is closed under contexts and substitutions,
and a compatible well-founded order $\succ$
which is closed under substitutions. Here compatibility means
the inclusion ${\succcurlyeq \cdot \succ \cdot \succcurlyeq} \subseteq {\succ}$.
Recall that any well-founded weakly monotone algebra $(\AA,\succcurlyeq)$
gives rise to a pair $(\geqord[\succcurlyeq]{\AA},\gord[\succ]{\AA})$ of
relations over terms. It is well known that the pair
$(\geqord[\succcurlyeq]{\AA},\gord[\succ]{\AA})$ forms
a reduction pair.
\begin{prop} \label{prop:redpair}
A TRS $\RS$ is terminating if and only if there exist
an argument filtering $\pi$ and a reduction pair $(\succcurlyeq,\succ)$ such
that ${\pi(\DP(\RS))} \subseteq {\succ}$ and ${\pi(\RS)} \subseteq {\succcurlyeq}$.
\qed
\end{prop}
We write $f \blacktriangleright_\RS g$ if there exists a rewrite rule $l \rew r
\in \RS$ such that $\rt(l)=f$ and $g$ is a defined function symbol in $\Fun(r)$.
For a set $\GS$ of defined function symbols we denote by $\RS \upharpoonright \GS$
the set of rewrite rules $l \rew r \in \RS$ with $\rt(l)\in\GS$. The set $\U_\RS(t)$
of usable rules of a term $t$ is defined as $\RS\upharpoonright\{g \mid f
\blacktriangleright_\RS^\ast g \text{ for some defined function symbol $f$ in
$\Fun(t)$}\}$. Finally, if $\C$ is a set of dependency pairs then
$\U_\RS(\C)\defsym\bigcup_{l\rew r\in\C}\U_\RS(r)$.
We write $\U(\C)$ instead of $\U_\RS(\C)$ if $\RS$ is clear from the context.
We use $\Ce$ to denote the two rules $\mcs(x,y)\rew x$ and
$\mcs(x,y)\rew y$ for some fresh binary function symbol $\mcs$.
\begin{prop}[\cite{GTSF06,HM07}] \label{prop:ur}
Let $\RS$ be a TRS. If there exist an argument filtering $\pi$ and a reduction
pair $(\succcurlyeq,\succ)$ such that ${\pi(\DP(\RS))} \subseteq {\succ}$ and
${\pi(\U(\DP(\RS)))\cup\Ce} \subseteq {\succcurlyeq}$, then $\RS$ is terminating.
\qed
\end{prop}

The \emph{dependency graph} of a TRS $\RS$ (denoted by $\DG(\RS)$)
is a graph whose nodes are the dependency pairs of $\RS$. It contains
an edge from $s\rew t$ to $u\rew v$ whenever there exist substitutions
$\sigma$ and $\tau$ such that $t\sigma\rssrew{\RS}u\tau$. A \emph{strongly
connected component} (\emph{SCC} for short) of $\DG(\RS)$ is a maximal subset of
nodes such that for each pair of nodes $s\rew t$, $u\rew v$, there
exists a path from $s\rew t$ to $u\rew v$. We call an SCC \emph{trivial}
if it consists of a single node $s\rew t$ such that the only path from
that node to itself is the empty path. All other SCCs are called
\emph{nontrivial}.

\begin{prop} \label{prop:dg}
A TRS $\RS$ is terminating if and only if for every nontrivial SCC $\C$ in $\DG(\RS)$
there exist an argument filtering $\pi$ and a reduction pair $(\succcurlyeq,\succ)$
such that ${\pi(\C)} \subseteq {\succ}$ and ${\pi(\RS)} \subseteq {\succcurlyeq}$.
\qed
\end{prop}

\section{Progenitor and Progeny}
\label{sec:ancestors}

In this and the next section we show that for the basic
dependency pair method (potentially using argument
filterings) the induced derivational complexity is triple
exponentially bounded in the derivational complexity induced by
the base technique employed.

Before proceeding into the technical construction, we outline the proof plan.
We aim to bound the length of derivations in a given TRS. Since any
derivation in a terminating TRS is non-cycling, the \emph{length} of any derivation
is bounded exponentially in the \emph{size} of the occurring terms. On the other hand, the
\emph{size} of any term is bounded exponentially in its \emph{depth}.
Thus it suffices to show that the \emph{depth} of any term occurring
in a derivation is exponentially bounded in
the number of admitted dependency pair steps, which in turn is bounded by the
derivational complexity induced by the base technique employed.

In the proof, we introduce the \emph{progeny}
relation (see Definition~\ref{def:ancestorstep}), which is an extension of the descendant 
relation~\cite[Chapter~4]{Terese}. We use the progeny relation in order to extract derivations over
$\DP(\RS)\cup\RS$ from a given derivation over a TRS $\RS$ (see Definition~\ref{def:constructeddp}).
In Definition~\ref{def:ancestorgraph} we exploit this notion to define the \emph{progenitor graph},
which constitutes a suitable restriction of the progeny relation for a given derivation $A$.
The intuition behind progenitor graphs
is to define a graph that captures the dependency pair steps of the
$\DP(\RS)\cup\RS$-derivations extracted from $A$.
Moreover the graph is constructed such that
its size linearly bounds the height of the last term in $A$ and
the height of its components is bounded
by the number of admitted dependency pair steps.

For the remainder of this paper, let $\RS$ be a TRS.  We recall
the definition of descendants.
Let $A\colon s \completerew{p'}{l\rew r} t$ be a rewriting step,
and let $p\in\Pos(s)$. Then the \emph{descendants of $p$ in $t$}
(denoted by $p\backslash A$) are defined as follows:
\begin{equation*}
p\backslash A\defsym
\begin{cases}
\{p\} & \text{if } p<p' \text{ or } p\parallel p',\\
\{p'q_3q_2 \mid \atpos{r}{q_3} = \atpos{l}{q_1}\} & \text{if } p=p'q_1q_2 \text{ with } q_1\in\VPos(l),\\
\varnothing & \text{otherwise}
\tpkt
\end{cases}
\end{equation*}\smallskip

\noindent We also want to keep track of redex positions, not just of positions
in the context or the substitution of the rewrite rule.
This intuition is cast into the following definition.
\begin{defi}
\label{def:ancestorstep}
Let $A\colon s\completerew{p'}{l\rew r} t$ be a rewriting step,
and let $p\in\Pos(s)$. Then the \emph{progenies of $p$ in $t$}
(denoted by $\child{p}{A}$) are:
\begin{equation*}
\child{p}{A}\defsym
\begin{cases}
\{p\}
& \text{if } p< p' \text{ or } p\parallel p',\\
\{p'q_3q_2 \mid \atpos{r}{q_3} = \atpos{l}{q_1}\}
& \text{if } p= p' q_1q_2 \text{ with } q_1\in\VPos(l),\\
\{p'q_2 \mid \atpos{r}{q_2} = \atpos{l}{q_1}\}
& \text{if } p= p' q_1 \text{ with } q_1\in\FPos(l) - \{\epsilon\},\\
\{pq_1 \mid {{\atpos{r}{q_1}} \nprsubterm {l}} \land {q_1\in\FPos(r)}\}
& \text{if } p=p'
\tpkt
\end{cases}
\end{equation*}
If $q \in \child{p}{A}$, then we also say that
$p$ is a \emph{progenitor of $q$ in $s$}. We denote the set of progenitors
of $q$ in $s$ by $\anc{q}{A}$, i.e., we have $q\in\child{p}{A}$ if and only if $p\in\anc{q}{A}$.
For a set $P\subseteq\Pos(s)$, we define
$\child{P}{A}\defsym\bigcup_{p\in P}\child{p}{A}$.
\end{defi}

\begin{rem}
Note that the distinction between the last two cases corresponds
to the exclusion of rules
$l^\sharp \to u^\sharp$ from $\DP(\RS)$ where $u \prsubterm l$, 
see Section~\ref{sec:DPMethod}. If we were not considering
the exclusion of those rules, we could omit the third case in
Definition~\ref{def:ancestorstep}, and drop the condition
$\atpos{r}{q_1}\nprsubterm l$ from the last case.
\end{rem}

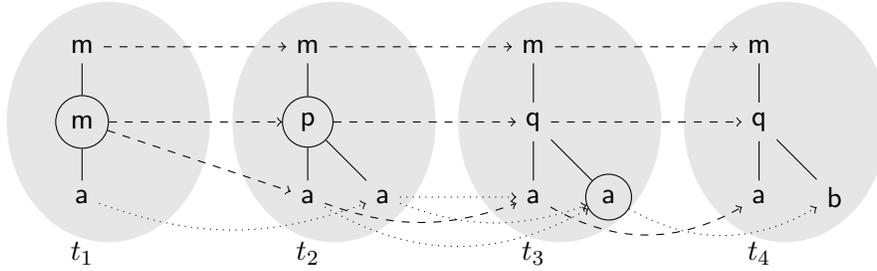
\begin{figure}[t]
\begin{center}
\begin{tikzpicture}
\fill[black!10!white] (0.35,0) ellipse (1.35cm and 1.60cm);
\fill[black!10!white] (3.35,0) ellipse (1.35cm and 1.60cm);
\fill[black!10!white] (6.35,0) ellipse (1.35cm and 1.60cm);
\fill[black!10!white] (9.35,0) ellipse (1.35cm and 1.60cm);

\node[circle, draw] (1te) {$\mm$};
\node[below of=1te] (1t1) {$\ma$};
\node[above of=1te] (1tc) {$\mm$};
\draw (1te) -- (1t1);
\draw (1tc) -- (1te);

\node[below of=1te, node distance=1.75cm] (1tcaption) {$t_1$};

\node[right of=1te, node distance=3cm, circle, draw] (2te) {$\mpp$};
\node[below of=2te] (2t1) {$\ma$};
\node[right of=2t1] (2t2) {$\ma$};
\node[above of=2te] (2tc) {$\mm$};
\draw (2te) -- (2t1);
\draw (2te) -- (2t2);
\draw (2tc) -- (2te);

\node[below of=2te, node distance=1.75cm] (2tcaption) {$t_2$};

\node[right of=2te, node distance=3cm] (3te) {$\mq$};
\node[below of=3te] (3t1) {$\ma$};
\node[circle, draw, right of=3t1] (3t2) {$\ma$};
\node[above of=3te] (3tc) {$\mm$};
\draw (3te) -- (3t1);
\draw (3te) -- (3t2);
\draw (3tc) -- (3te);

\node[below of=3te, node distance=1.75cm] (3tcaption) {$t_3$};

\node[right of=3te, node distance=3cm] (4te) {$\mq$};
\node[below of=4te] (4t1) {$\ma$};
\node[right of=4t1] (4t2) {$\mb$};
\node[above of=4te] (4tc) {$\mm$};
\draw (4te) -- (4t1);
\draw (4te) -- (4t2);
\draw (4tc) -- (4te);

\node[below of=4te, node distance=1.75cm] (4tcaption) {$t_4$};

\draw[dashed,->] (1te) to (2te);
\draw[dashed,->] (1te) to (2t1);
\draw[dotted,->,bend right=20] (1t1) to (2t2);
\draw[dashed,->] (1tc) to (2tc);

\draw[dashed,->] (2te) to (3te);
\draw[dashed,->,bend right=20] (2t1) to (3t1);
\draw[dotted,->,bend right=30] (2t1) to (3t2);
\draw[dotted,->] (2t2) to (3t1);
\draw[dotted,->,bend right=20] (2t2) to (3t2);
\draw[dashed,->] (2tc) to (3tc);

\draw[dashed,->] (3te) to (4te);
\draw[dashed,->,bend right=30] (3t1) to (4t1);
\draw[dotted,->,bend right=30] (3t2) to (4t2);
\draw[dashed,->] (3tc) to (4tc);
\end{tikzpicture}
\end{center}
\caption{A Derivation, its Progeny Relation and Redex Positions}
\label{fig:ancestors}
\end{figure}

\begin{exa}
\label{ex:ancestors}
Consider the TRS $\RSb$ consisting of the following three rewrite rules:
\begin{equation*}
\mm(x)\rew\mpp(\ma,x) \qquad
\mpp(x,x)\rew\mq(x,x) \qquad
\ma\rew\mb
\tpkt
\end{equation*}
Let $A$ be the derivation
\begin{equation*}
\underbrace{\mm(\mm(\ma))}_{t_1}\rew
\underbrace{\mm(\mpp(\ma,\ma))}_{t_2}\rew
\underbrace{\mm(\mq(\ma,\ma))}_{t_3}\rew
\underbrace{\mm(\mq(\ma,\mb))}_{t_4}
\tkom
\end{equation*}
which is represented in Figure~\ref{fig:ancestors}. Redex positions
are marked by circles, the progeny relation is marked by dotted and dashed
lines (the two kinds of lines will be distinguished in Example~\ref{ex:mainancestors} below).
Note that each position in a term may have several progenitors. For instance,
$\anc{11}{(t_2 \rew t_3)}=\{11,12\}$.
\end{exa}

\begin{lem}
\label{lem:ancestortodp}
Let $A \colon s \rew t$, let
$p\in\Pos(s)$, and let $q\in\Pos(t)$.
If $q \in \child{p}{A}$ and $\rt(\atpos{t}{q})\in\DS$,
then  $\rt(\atpos{s}{p})\in\DS$ and $(\atpos{s}{p})^\sharp  \rsgrew{\DP(\RS) \cup \RS} (\atpos{t}{q})^\sharp$.
\end{lem}
\proof
Suppose that $A$ is $s \completerew{p'}{l\rew r} t$.
If $p < p'$ or $p \parallel p'$, then by definition, we have
$p = q$ and thus $(\atpos{s}{p})^\sharp \rsgrew{\RS} (\atpos{t}{q})^\sharp$.
On the other hand, if $p=p'$, then there exists $q_1 \in \FPos(r)$ such that
$q=p'q_1$. Moreover, ${\atpos{t}{q}} \nprsubterm {\atpos{s}{p}}$. By assumption
$\rt(\atpos{t}{q})\in\DS$ and thus we obtain
$(\atpos{s}{p})^\sharp \rsrew{\DP(\RS)} (\atpos{t}{q})^\sharp$.
Finally, if $p > p'$, then by definition of the progeny relation, we have $\atpos{s}{p} = \atpos{t}{q}$. Then
again, $(\atpos{s}{p})^\sharp \rsgrew{\RS} (\atpos{t}{q})^\sharp$ follows trivially.
\qed

\begin{lem}
\label{lem:allhasancestorstep}
Let $A \colon s \rew t$. Then for every $q\in\Pos(t)$,
we have $\anc{q}{A}\neq\varnothing$.
\end{lem}
\proof
Suppose $A$ denotes the step $s\completerew{p'}{l\rew r} t$.
If $q<p'$ or $q\parallel p'$, then $\anc{q}{A}=\{q\}$. If $q=p'q_1$,
$q_1\in\FPos(r)$, and $\atpos{r}{q_1} \nprsubterm l$, then $\anc{q}{A}=\{p'\}$.
If $q=p'q_1$, $q_1\in\FPos(r)$, and $\atpos{r}{q_1} \prsubterm l$, then
there is some $p_1$ such that $\atpos{l}{p_1}=\atpos{r}{q_1}$, so $p'p_1\in \anc{q}{A}$. Last,
if $q=p'q_1q_2$ and $q_1\in\VPos(r)$, then there is some $p_1$ such that
$\atpos{l}{p_1}=\atpos{r}{q_1}$ because $\Var(r)\subseteq\Var(l)$.
Therefore, $p'p_1q_2\in \anc{q}{A}$.
\qed

\begin{defi}
\label{def:ancestorseq}
Let $A\colon s\rss t$ be a derivation, and let $p\in\Pos(s)$.
Then the \emph{progenies of $p$ in $t$} (also denoted by $\child{p}{A}$) are
defined as follows:
\begin{enumerate}[(1)]
\item If $A$ is the empty derivation, then $\child{p}{A}=\{p\}$.
\item Otherwise, we can split $A$ into $A_1:s\rew s'$ and $A_2:s'\rss t$.
Then $\child{p}{A}=\child{(\child{p}{A_1})}{A_2}$.
\end{enumerate}
We say $p$ is a \emph{progenitor} of $q$ if $p\in \anc{q}{A}$, which
holds if $q\in \child{p}{A}$. Moreover, we have
$q\in \child{P}{A}$ if and only if $q\in \child{p}{A}$ for some $p\in P$.
\end{defi}

\begin{lem}
Let $A \colon s \rss t$ be a derivation and 
let $p\in\Pos(s)$, $q\in\Pos(t)$.
\label{lem:allhasancestor}
\label{lem:ancestortodpseq}
Then the set $\anc{q}{A}$ of progenitors of $q$ is not empty. 
Moreover if $q\in \child{p}{A}$ with $\rt(\atpos{t}{q})\in\DS$,
then $\rt(\atpos{s}{p})\in\DS$ and $(\atpos{s}{p})^\sharp \rssrew{\DP(\RS) \cup \RS} (\atpos{t}{q})^\sharp$.
\end{lem}
\proof
Straightforward induction using
Lemmata~\ref{lem:allhasancestorstep} and~\ref{lem:ancestortodp}.
\qed

Using Lemma~\ref{lem:ancestortodpseq}, we can extract derivations over $\DP(\RS)\cup\RS$
from a given derivation in a TRS $\RS$ using positions connected by the progeny relation.
\begin{defi}
\label{def:constructeddp}
Let $\seq{t}$ be terms, and let $\seq{p}$ be positions
in $\seq{t}$, respectively, such that $\rt(\atpos{t_n}{p_n})\in\DS$, and for all
$1\leqslant i\leqslant n-1$, we have $A_i \colon t_i\rsrew{\RS} t_{i+1}$ and
$p_{i+1}\in \child{p_i}{A_i}$. Then we call
$A\colon (\atpos{t_1}{p_1})^\sharp\rssrew{\DP(\RS)\cup\RS}(\atpos{t_n}{p_n})^\sharp$
the \emph{implicit dependency pair derivation} with respect to $\seq{t}$ and $\seq{p}$.
We denote the number of $\DP(\RS)$-steps in $A$ as $\dpsize{A}$.
\end{defi}

\begin{exa}[continued from Example~\ref{ex:ancestors}]
The implicit dependency pair derivation with respect to the terms $t_1$, $t_2$, $t_3$ and
the positions $1$, $11$, $12$ is given as follows:
\begin{equation*}
  \mm^\sharp(\ma)\;\rsrew{\DP(\RSb)}\;
  \ma^\sharp\;\rsgrew{\RSb}\;
  \ma^\sharp
  \tpkt
\end{equation*}
Note that the length of this implicit dependency pair derivation is smaller
than the length of the original derivation $A$. Moreover, all terms occurring
in this implicit dependency pair derivation are proper subterms of the
respective terms of $A$ (modulo marking top symbols by a $\sharp$). In contrast,
the implicit dependency pair derivation with respect to the terms $t_1$, $t_2$, $t_3$,
$t_4$, and the positions $\epsilon$, $\epsilon$, $\epsilon$, $\epsilon$ is given by
$t_1^\sharp\rsrew{\RSb}t_2^\sharp \rsrew{\RSb} t_3^\sharp\rsrew{\RSb}t_4^\sharp$.
\end{exa}

\begin{figure}
\begin{center}
\begin{tikzpicture}
\node[above] (captionS) at (0,0) {$s$};
\node[below, regular polygon, regular polygon sides=3, minimum size=3cm, draw] (termS) at (0,0) {};

\node[above] (captionT) at (7,0) {$t$};
\node[below, regular polygon, regular polygon sides=3, minimum size=3cm, draw] (termT) at (7,0) {};

\node[above, regular polygon, regular polygon sides=3, minimum size=1.2cm, draw] (posP) at (termT.south) {};
\node[right] (captionP) at (posP.north) {$q$};

\node[above, regular polygon, regular polygon sides=3, minimum size=0.33cm, inner sep=0cm, draw] (posQ) at (posP.south) {};
\node[right] (captionQ) at (posQ.north) {$q'$};

\node[above, xshift=-0.6cm, regular polygon, regular polygon sides=3, minimum size=1.2cm, draw] (posP2) at (termS.south) {};
\node[left] (captionP2) at (posP2.north) {$p_0$};

\node[above, regular polygon, regular polygon sides=3, minimum size=0.33cm, inner sep=0cm, draw] (posQ2) at (posP2.south) {};
\node[left] (captionQ2) at (posQ2.north) {$p_0'$};

\node[above, xshift=0.6cm, regular polygon, regular polygon sides=3, minimum size=1.2cm, draw] (posP3) at (termS.south) {};
\node[left] (captionP3) at (posP3.north) {$p_1$};

\node[above, regular polygon, regular polygon sides=3, minimum size=0.33cm, inner sep=0cm, draw] (posQ3) at (posP3.south) {};
\node[left] (captionQ3) at (posQ3.north) {$p_1'$};

\draw[->, bend right=15] (posP.north) to (posP2.north);
\draw[->, bend right=10] (posP.north) to (posP3.north);

\draw[->] (posP.north) to (posQ.north);

\draw[->, dotted] (posP2.north) to (posQ2.north);
\draw[->, dotted] (posP3.north) to (posQ3.north);

\draw[->, dotted, bend left=15] (posQ.north) to (posQ2.north);
\draw[->, dotted, bend left=10] (posQ.north) to (posQ3.north);
\end{tikzpicture}
\end{center}
\caption{Intuition for Lemma \ref{lem:relativepositions}}
\label{fig:relativepositions}
\end{figure}
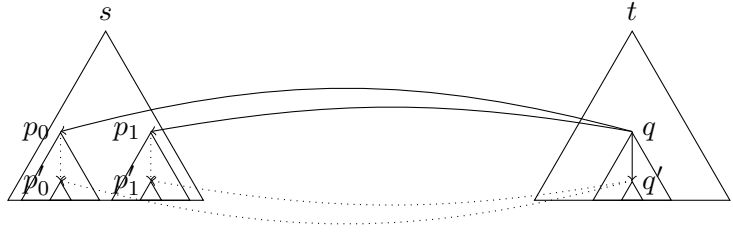

The following lemma shows that given two positions $q\leqslant q'$ in the same branch of a term,
and a progenitor $p_0$ of $q$, we can always find a progenitor
$p_0'$ of $q'$ such that $p_0\leqslant p_0'$.  
This is graphically depicted in Figure~\ref{fig:relativepositions}, where the
drawn lines indicate the assumption of the lemma and the dotted lines the
conclusion.
The lemma entails that for any branch $B$ of a term,
we can find progenitors of all positions in $B$ in a single branch again.

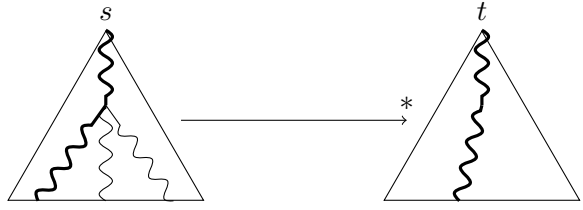
\begin{figure}
\begin{center}
\begin{tikzpicture}
\node[above] (captionS) at (0,0) {$s$};
\node[below, regular polygon, regular polygon sides=3, minimum size=3cm, draw] (termS) at (0,0) {};
\node[above] (captionT) at (5,0) {$t$};
\node[below, regular polygon, regular polygon sides=3, minimum size=3cm, draw] (termT) at (5,0) {};
\draw[->] (1,-1.2) -> (4,-1.2) node[above] {$*$};
\draw[decorate, decoration={coil, aspect=0}, style=very thick] (termS.north) -- (0,-1.0) node[scale=0.01] (mycenterS) {};
\draw[decorate, decoration={coil, aspect=0}, style=very thick] (termS.south) ++(-0.9,0) -- (mycenterS);
\draw[decorate, decoration={coil, aspect=0}] (termS.south) -- (mycenterS);
\draw[decorate, decoration={coil, aspect=0}] (termS.south) ++(0.9,0) -- (mycenterS);
\draw[decorate, decoration={coil, aspect=0}, style=very thick] (termT.north) -- (5,-1.0) node[scale=0.01] (mycenterT) {};
\draw[decorate, decoration={coil, aspect=0}, style=very thick] (termT.south) ++(-0.3,0) -- (mycenterT);
\end{tikzpicture}
\end{center}
\caption{Depiction of the Main Branch in a Derivation}
\label{fig:mainbranch}
\end{figure}

\begin{lem}
\label{lem:relativepositions}
Let $A \colon s \rew t$ and let $q,q' \in\Pos(t)$.
If $q \leqslant q'$, then for any $p_0 \in \anc{q}{A}$, there exists
$p_0'\in \anc{q'}{A}$ such that $p_0 \leqslant p_0'$.
\end{lem}
\proof
Suppose $A$ has the form $s\completerew{p'}{l\rew r} t$.
According to Definition~\ref{def:ancestorstep}, there are four cases for $q'$.
\begin{enumerate}[(1)]
\item If $q'<p'$ or $q'\parallel p'$, then also $q<p'$ or
$q\parallel p'$.
Therefore, $\anc{q}{A}=\{q\}$ and $\anc{q'}{A}=\{q'\}$.
\item If $q'=p'q_1'$, $q_1'\in\FPos(r)$,
and $\atpos{r}{q_1'}\nprsubterm l$, then either $q<p'$,
or $q=p'q_1$, $q_1\in\FPos(r)$, and $\atpos{r}{q_1}\nprsubterm l$.
We have $\anc{q}{A}=\{p_0\}$ and $\anc{q'}{A}=\{p'\}$ with $p_0=q$ or $p_0=p'$.
In both cases, $p_0\leqslant p'$, so the lemma follows.
\item If $q'=p'q_1'$, $q_1'\in\FPos(r)$, and $\atpos{r}{q_1'}\prsubterm l$,
then $\anc{q'}{A}=\{p'q_2'\mid q_2'\in\FPos(l)\land \atpos{r}{q_1'}=\atpos{l}{q_2'}\}$.
From the three cases in Definition~\ref{def:ancestorstep} applicable
for $q$, we only consider the last one, where $q = p'q_1$, $q_1 \in \FPos(r)$
and $\atpos{r}{q_1}\prsubterm l$,
then $\anc{q}{A}=\{p'q_2\mid q_2\in\FPos(l)\land \atpos{r}{q_1}=\atpos{l}{q_2}\}$. Since
$q\leqslant q'$, there exists some $q_3'$ such that $q'=qq_3'$. Hence, for any
$p'q_2\in \anc{q}{A}$, we also have $p'q_2q_3'\in \anc{q'}{A}$, entailing the lemma.

\item If $q'=p'q_1'q_2'$ with $q_1'\in\VPos(r)$, then
$\anc{q'}{A}=\{p'q_3'q_2'\mid \atpos{r}{q_1'}=\atpos{l}{q_3'}\}$. Except for $q\parallel p'$,
all cases are possible for $q$. 
Again, we restrict to one of these
cases and assume that $q=p'q_1'q_2$. Then
$\anc{q}{A}=\{p'q_3q_2\mid \atpos{r}{q_1'}=\atpos{l}{q_3}\}$.
Since $q\leqslant q'$, there exists $q_4'$ such that $q_2'=q_2q_4'$.
Hence, for any $p'q_3q_2\in \anc{q}{A}$, we also have $p'q_3q_2'\in \anc{q'}{A}$,
thus the lemma follows.
\qed
\end{enumerate}
In order to simplify the structure of the progeny relation we restrict the progenies
and progenitors to a single branch in each term.
The definition rests on the idea that for a derivation $A\colon s \rss t$ and a
\emph{main branch} $B'$ in $t$ it is possible to
find a main branch $B$ in $s$ such that each position $q \in B'$ has a (unique)
progenitor in $B$. See Figure~\ref{fig:mainbranch} for an illustration. The bold
lines denote the main branches of $s$ and $t$, and the thin lines denote other
branches of $s$ containing progenitors of all positions in the main branch of $t$.

\begin{defi}
\label{def:mainbranch}
Let $A\colon t_1 \rss t_n$ denote a derivation built
up from the rewrite steps $A_i\colon t_i \rew t_{i+1}$ for $i= 1,\dots,n-1$.
Then the \emph{main branch} of each term in $A$ is inductively defined:
\begin{enumerate}[(1)]
\item The main branch of $t_n$ is the leftmost branch among all branches of maximal length in $t_n$.
\item Suppose the main branch of $t_{i+1}$ is denoted as $B_{i+1}$, $1 \leqslant i \leqslant n-1$.
Then consider all branches $b$ in $t_i$ such that for every $q \in B_{i+1}$, the set of progenitors
$\anc{q}{A_i}$ of $q$ has nonempty intersection with $b$. The leftmost of these branches
is the main branch of $t_i$, denoted as $B_i$.
\end{enumerate}
\end{defi}
\noindent
In the above definition, the restriction to the leftmost of all candidate branches
is arbitrary and can be suitably replaced. The second clause is well-defined
by Lemmata~\ref{lem:allhasancestor} and~\ref{lem:relativepositions}. Note that a branch
of maximal size is chosen for the final term of the given derivation since it
reflects the depth of this term, c.f.~Section~\ref{sec:dpcomplexity}.
The next definition specialises progenies and progenitors to the main branch.
\begin{defi}
\label{def:mainancestor}
Let $A':s\rew t$ be a rewriting step, let $p\in\Pos(s)$, and let $B$ and $B'$
be branches in $s$ and $t$, respectively. Then the set of \emph{main progenies of $p$ in $t$}
(with respect to $A'$) (denoted as $\mchild{p}{A'}{B}{B'}$) is defined as follows:
\begin{equation*}
\mchild{p}{A'}{B}{B'}\defsym
\begin{cases}
\varnothing & \qquad\text{if }p\notin B\\
B'\cap (\child{p}{A'}) & \qquad\text{if }p\in B
\tpkt
\end{cases}
\end{equation*}

If the (main) branches $B$ and $B'$ are clear from context, we write $\mchildB{p}{A'}$
instead of $\mchild{p}{A'}{B}{B'}$.
If $q\in \mchildB{p}{A'}$, then we also say that
$p$ is a \emph{main progenitor of $q$ in $s$} (with respect to $A'$).
We denote the set of main progenitors of $q$ in $s$
by $\mancB{q}{A'}$.
For a set $P \subseteq\Pos(s')$, we define
$\mchildB{P}{A'}\defsym\bigcup_{p\in P}\mchildB{p}{A'}$.
We naturally extend the definition to derivations $A\colon s\rss t$, analogous to
Definition~\ref{def:ancestorseq}:
if $A$ is the empty derivation, then $\mchild{p}{A}{B}{B'}=\{p\}$.
Otherwise, we can split $A$ into $A_1:s\rew s'$ and $A_2:s'\rss t$.
Let $B''$ be the main branch in $s'$.
Then $\mchild{p}{A}{B}{B'}=\mchild{(\mchild{p}{A_1}{B}{B''})}{A_2}{B''}{B'}$.
\end{defi}

\begin{lem}
\label{lem:uniquemainancestor}
Let $A\colon u\rss s \rew^n t\rss w$ and denote $A'\colon s \rew^n t$.
Let $B(s)$ and $B(t)$ denote the main branches of $s$ and $t$ in $A$, respectively.
Then for any $q \in B(t)$, the main progenitor
of $q$ in the branch $B(s)$ is unique, i.e., $\card{\mancB{q}{A'}}=1$.
\end{lem}
\proof
By Definition~\ref{def:mainbranch}, $q$ has at least one main progenitor in $s$. We show that there
exists at most one by induction on $n$.
For $n=0$ the claim is trivial. Hence assume $n>0$ and let $A' \colon s \rew s' \rew^{n-1} t$.
Let $B(s')$ denote the main branch in $s'$ with respect to $A$. By induction hypothesis
there exists a unique position $p_1$ in $B(s')$ such that $\mancB{q}{(s' \rew^{n-1} t)} = \{p_1\}$.
Let $A'' \colon s \completerew{p'}{l \rew r} s'$ denote the first rewrite step in $A'$.
Suppose $p_1 < p'$ or $p_1 \parallel p'$. Then by definition $\anc{p_1}{A''} = \{p_1\}$.
Hence the main progenitor of $q$ in $B(s)$ is unique. On the other
hand suppose $p_1 = p' p_2$ with $p_2 \in \FPos(r)$ such that $\atpos{r}{p_2} \nprsubterm l$.
Then $\anc{p_1}{A''} = \{p'\}$ and $\mancB{q}{A'}$ is a singleton as it should be.
Now suppose  $p_1 = p' p_2$ with $p_2 \in \FPos(r)$ such that $\atpos{r}{p_2} \prsubterm l$.
Then by definition $\anc{p_1}{A''} = \{p' p_3 \mid p_3 \in \FPos(l) \land \atpos{l}{p_3} = \atpos{r}{p_2}\}$.
Note that $\mancB{q}{A'} = \anc{p_1}{A''} \cap B(s)$, which is again a singleton. Finally,
if $p_1 = p' p_2 p_3$ with $p_2 \in \VPos(r)$, then
$\anc{p_1}{A''} = \{p' p_4 p_3 \mid p_4 \in \VPos(l) \land \atpos{l}{p_4} = \atpos{r}{p_2}\}$.
As before, the intersection of the latter set with $B(s)$ is a singleton.
Hence the main progenitor of $q$ in $B(s)$ is unique.
This concludes the inductive proof.
\qed

Observe that we cannot define main progenies for (multi-step) derivations directly by restricting the
progeny relation to the main branches; it is indeed necessary to use the inductive definition
given above. In particular, Lemma~\ref{lem:uniquemainancestor} would be incorrect for
that definition, as exemplified below.
\begin{exa}[continued from Example~\ref{ex:ancestors}]
\label{ex:mainancestors}
Consider the derivation $A$ again. We split $A$ into
$A_1:t_1\rew t_2$, $A_2:t_2\rew t_3$, and $A_3:t_3\rew t_4$. The ``central'' branch of each term
in Figure~\ref{fig:ancestors} is its main branch, and the dashed lines
denote the main progeny relation. Note that $1\in\mancB{11}{A}$,
since $11\in\mancB{11}{A_3}$, $11\in\mancB{11}{A_2}$, and $1\in\mancB{11}{A_1}$. Furthermore, we do not have
$11\in\mancB{11}{A}$, even though $11\in\anc{11}{A_3}$, $12\in\anc{11}{A_2}$, $11\in\anc{12}{A_1}$,
and therefore $11\in\anc{11}{A}$.
\end{exa}

For positions pointing to non-defined symbols, we also have the reverse of
Lemma~\ref{lem:uniquemainancestor}.
\begin{lem}
\label{lem:noconstructorbranching}
We assume the same notation as in Lemma~\ref{lem:uniquemainancestor}.
For any $p\in B(s)$ such that $\rt(\atpos{s}{p})\in\CS\cup\VS$, we have
$\card{\mchildB{p}{A'}}\leqslant 1$, i.e., the number of main progenies
for a position such that the root of the corresponding subterm is non-defined is at most~$1$.
\end{lem}
\proof
By induction on $n$.
It suffices to consider the case $n>0$, so
$A' \colon s \rew s' \rew^{n-1} t$. Let $A'' \colon s \completerew{p'}{l \rew r} s'$
denote the first rewrite step in $A'$.
If $p<p'$ or $p\parallel p'$, then $\mchildB{p}{A''}=\{p\}$.
If $p>p'$, then for any $p_1 \in\child{p}{A''}$, we have
$\atpos{s}{p} = \atpos{s'}{p_1}$, so again, $\child{p}{A''}\cap B(s')$ is a singleton.
In all of these cases, the claim follows by induction hypothesis as
$\rt(\atpos{s}{p})=\rt(\atpos{s'}{p_1})$ for any $p_1\in\mchildB{p}{A''}$. This concludes the
proof, as the case $p=p'$ is impossible. Otherwise, we derive a contradiction
to the assumption that the root of $\atpos{s}{p}$ is not a defined symbol.
\qed

\section{Dependency Pairs and Complexity}
\label{sec:dpcomplexity}

Let $A \colon t_1 \rssrew{\RS} t_n$ be a derivation with respect to $\RS$,
and let $m$ be the maximum number of $\DP(\RS)$-steps in any implicit
dependency pair derivation corresponding to $A$.
In this section we show that the length $n$ of $A$ is
bounded triple exponentially in $m$.
As mentioned at the beginning of Section~\ref{sec:ancestors}, it suffices to
show that the depth of any term occurring in $A$ is exponentially bounded in $m$.
More precisely, as we consider an arbitrary derivation $A$, it even suffices to
show that the depth of the term $t_n$ is exponentially bounded in $m$,
c.f.~Lemma~\ref{lem:dp2height}.

\begin{notation}
In the sequel, we fix the derivation $A$ and let $B_1$, \dots, $B_n$
denote the main branches of $t_1$, \dots, $t_n$ with respect to~$A$.
Let $G$ be the progenitor graph of $A$ (see
Definition~\ref{def:ancestorgraph} below).
For the remainder of this paper, let
$C \defsym \max(\{2\}\cup\{\depth{r}+1\mid l\rew r\in\RS\})$. We call
$C$ the \emph{branching constant} of $\RS$.
\end{notation}

In the next definition we formalise \emph{progenitor graphs}.
\begin{defi}
\label{def:ancestorgraph}
The \emph{progenitor graph} $G$ of $A$ is defined as follows.
\begin{enumerate}[(1)]
\item The nodes are all pairs $(t_i,p)$ such that $p\in B_i$ with
$\rt(\atpos{t_i}{p})$ defined and either $i=1$ 
or the single element of $\mancB{p}{(t_{i-1}\rew t_i)}$ and
the redex position in the rewrite step $t_{i-1}\rew t_i$ coincide.
\item There is an edge from $(t_i,p)$ to $(t_j,q)$ whenever $i<j$,
$\mancB{q}{(t_i\rss t_j)}=\{p\}$, and for all $i\leqslant k<j-1$, the
single element of $\mancB{q}{(t_k\rss t_j)}$ and the redex position in
the rewrite step $t_k\rew t_{k+1}$ do \emph{not} coincide.
\end{enumerate}
With respect to the definition of edges note that 
the single element of $\mancB{q}{(t_{j-1}\rew t_j)}$
and the redex position in the rewrite step $t_{j-1}\rew t_j$ coincide.
Also note that $G$ is a forest, and the root of each tree in $G$ is
$(t_1,p)$ for some $p\in B_1$.
\end{defi}

\begin{figure}
\begin{center}
\begin{tikzpicture}
\node (1te) {$(t_1,1)$};
\node[below of=1te] (1t1) {$(t_1,11)$};
\node[above of=1te] (1tc) {$(t_1,\epsilon)$};
\node[right of=1te, node distance=3cm] (2te) {$(t_2,1)$};
\node[below of=2te] (2t1) {$(t_2,11)$};
\draw[->] (1te) to (2te);
\draw[->] (1te) to (2t1);
\end{tikzpicture}
\end{center}
\caption{Progenitor Graph}
\label{fig:progenitorgraph}
\end{figure}
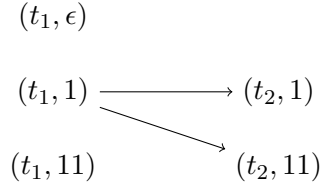

\begin{exa}
\label{ex:agraphsimp}
Consider the derivation $A$ from Example~\ref{ex:ancestors} again.
Its progenitor graph $G$ is shown in Figure~\ref{fig:progenitorgraph}.
Observe that $(t_2,\epsilon)$ is not contained in $G$ since the single
element of $\mancB{\epsilon}{(t_1\rew t_2)}$ is $\epsilon$, and $\epsilon$
is not the redex position of the step $t_1\rew t_2$.
For similar reasons, $(t_3,\epsilon)$, $(t_3,11)$, $(t_4,\epsilon)$,
$(t_4,1)$, and $(t_4,11)$ are not contained in $G$.
Moreover $(t_3,1)$ is not contained in $G$ either because
$\rt(\atpos{t_3}{1})=\mq$ is not defined.
Furthermore, $(t_2,12)$, $(t_3,12)$, and $(t_4,12)$ are not contained in
$G$ because $12$ is not a member of the main branch of $t_2$, $t_3$, and
$t_4$, respectively.

However, $(t_1,\epsilon)$, $(t_1,1)$, and $(t_1,11)$ are still contained in
$G$ because all positions of $t_1$ pointing to defined symbols are in $G$.
Moreover $(t_2,1)$ is contained in $G$ because $\rt(\atpos{t_2}{1})\in\DS$,
and the single element of $\mancB{1}{(t_1\rew t_2)}$ is the redex position
of the step $t_1\rew t_2$. For the same reason, $(t_2,11)$ is contained in
$G$.
\end{exa}

The main factor of the exponentially faster growth of $\depth{t_n}$ compared to
the maximal height of all trees in $G$ is the difference between that maximal
height and the size of $G$ (which is linearly related to $\depth{t_n}$).
This becomes apparent in our next example, where $G$ is a full binary tree.

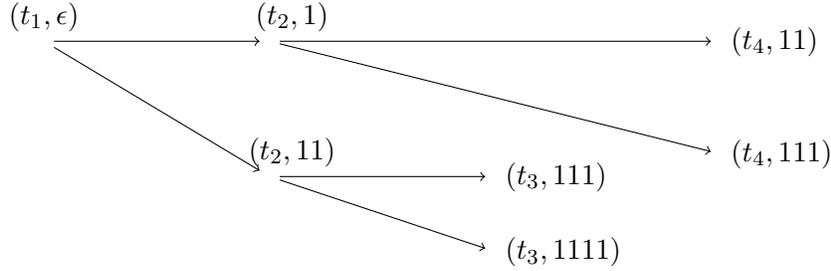
\begin{figure}
\begin{center}
\begin{tikzpicture}
\node (1te) {};
\node[below of=1te] (1t1) {};
\node[right of=1te, xshift=20mm] (2t1) {};
\node[below of=2t1, yshift=-8mm] (2t11) {};
\node[right of=2t11, xshift=20mm] (3t111) {};
\node[below of=3t111] (3t1111) {};
\node[right of=3t111, xshift=20mm, yshift=3mm] (4t111) {};
\node[above of=4t111, yshift=5mm] (4t11) {};

\draw (1te) node[above] {$(t_1,\epsilon)$};
\draw (2t1) node[above, xshift=3mm] {$(t_2,1)$};
\draw (2t11) node[above, xshift=3mm] {$(t_2,11)$};
\draw (3t111) node[right] {$(t_3,111)$};
\draw (3t1111) node[right] {$(t_3,1111)$};
\draw (4t11) node[right] {$(t_4,11)$};
\draw (4t111) node[right]{$(t_4,111)$};

\draw[->] (1te) -> (2t1);
\draw[->] (1te) -> (2t11);
\draw[->] (2t1) -> (4t11);
\draw[->] (2t1) -> (4t111);
\node[below of=2t11] (2t111) {};
\draw[->] (2t11) -> (3t111);
\draw[->] (2t11) -> (3t1111);
\end{tikzpicture}
\end{center}
\caption{Progenitor Graph: Full Binary Tree}
\label{fig:progenitorgraph:2}
\end{figure}

\begin{exa}
\label{ex:agraphprinciple}
Consider the TRS $\RSd \defsym \{\mfull(\ms(x))\rew\ms(\mfull(\mfull(x)))\}$
together with the following derivation $A$:
\begin{equation*}
\underbrace{\mfull(\ms(\ms(\Null)))}_{t_1}\rew
\underbrace{\ms(\mfull(\mfull(\ms(\Null))))}_{t_2}\rew
\underbrace{\ms(\mfull(\ms(\mfull(\mfull(\Null)))))}_{t_3}\rew
\underbrace{\ms(\ms(\mfull(\mfull(\mfull(\mfull(\Null))))))}_{t_4} \tpkt
\end{equation*}
The progenitor graph of $A$ is shown in Figure~\ref{fig:progenitorgraph:2}.
\end{exa}

\begin{lem}
\label{lemagf:edgederivation}
If there is an edge from $(t_i,p)$ to $(t_j,q)$ in $G$, then there exists
$q' \in B_{j-1}$ such that there is a derivation 
$(\atpos{t_i}{p})^\sharp\rssrew{\RS}(\atpos{t_{j-1}}{q'})^\sharp\rsrew{\DP(\RS)}(\atpos{t_j}{q})^\sharp$.
\end{lem}
\proof
By definition, $q\in\mchildB{p}{(t_i\rss t_j)}$.
Therefore, by Lemma~\ref{lem:ancestortodpseq}, we have the implicit dependency pair derivation
$A' \colon (\atpos{t_i}{p})^\sharp \rssrew{\DP(\RS) \cup \RS} (\atpos{t_j}{q})^\sharp$.
We have $\mancB{q}{(t_{j-1}\rew t_j)}=\{q'\}$, where by definition $q'$ is the redex position
of the step $t_{j-1}\rew t_j$. Therefore, the last step of $A'$ is a $\DP(\RS)$-step
(see also the last clause of Definition~\ref{def:ancestorstep}).
Note that for $i\leqslant k<j-1$, the single element of $\mancB{q}{(t_k\rss t_j)}$
and the redex position in $t_k\rew t_{k+1}$ do not coincide. Hence, if there are rewrite
steps before the last step, these are $\RS$-steps and the lemma follows.
\qed

The next lemma shows that Definition~\ref{def:ancestorgraph} is well-defined 
in the sense that only those nodes that do not contribute to the branching of the 
progenitor graph are left out.
\begin{lem}
\label{lem:hideRsteps}
Let $p\in B_i$ and $q\in B_j$ such that
$i<j$ and $\mancB{q}{(t_i\rss t_j)}=\{p\}$. If for all
$i\leqslant k\leqslant j-1$, the single element of $\mancB{q}{(t_k\rss t_j)}$
and the redex position in the rewrite step $t_k\rew t_{k+1}$ do not coincide,
then $\mchildB{p}{(t_i\rss t_j)}=\{q\}$.
\end{lem}
\proof
We show the lemma by induction on $j-i$.
If $i=j$ then the claim trivially
holds. Otherwise, the derivation $t_i\rss t_j$ can be split
into $t_i\rew t_{i+1}\rss t_j$. Let $p'$ be the redex position in $t_i\rew t_{i+1}$.
If $p\parallel p'$, $p<p'$, or $p>p'$, then as in
Lemma~\ref{lem:noconstructorbranching},
$\card{\mchildB{p}{(t_i\rew t_{i+1})}}\leqslant1$, and the lemma follows
by induction hypothesis. The remaining case is again impossible, since by
assumption, $p$ and $p'$ do not coincide.
\qed

In the following sequence of lemmata we show the properties which allow us to bound
$\depth{t_n}$ in the maximal height of all trees in $G$. First, we prove that almost each position in
$B_n$ is ``covered'' by a node in $G$. Next, we show that there exists a fixed
upper bound on the number of positions in $B_n$ each node in $G$ can cover,
and finally, we show that there is a fixed upper bound on the branching factor
of $G$, as well.

\begin{lem}
\label{lemagf:allfinalcovered}
Let $k \in \{1,\dots,n\}$. 
For every $q\in B_k$, there exists a unique $p\in B_1$ such that
either $\rt{(\atpos{t_1}{p})}\in\CS\cup\VS$ and 
$\mancB{q}{(t_1\rss t_k)}=\{p\}$, or there exists a unique 
node $(t_i,p_i)$ in $G$ where $q\in\mchildB{p_i}{(t_i\rss t_k)}$ and
for any direct successor node $(t_j,p_j)$ of $(t_i,p_i)$ in $G$, we have
$q\notin\mchildB{p_j}{(t_j\rss t_k)}$.
\end{lem}
\proof
By Lemma~\ref{lem:uniquemainancestor},
$\mancB{q}{(t_1\rss t_k)}=\{p\}$ for some $p\in B_1$. If
$\rt(\atpos{t_1}{p})\in\CS\cup\VS$, the first alternative of the lemma holds.
If $\rt(\atpos{t_1}{p})\in\DS$, then
$(t_1,p)\in G$. Therefore, there exists a maximal natural number $i$ such that
$(t_i,p_i)\in G$ and $q\in\mchildB{p_i}{(t_i\rss t_k)}$ for some $p_i\in B_i$,
so the second alternative of the lemma holds for $(t_i,p_i)$.
\qed

Lemma~\ref{lemagf:allfinalcovered} suggest the following definition.
\begin{defi}
\label{d:cover}
Let $k \in \{1,\dots,n\}$ and let $q\in B_k$. Suppose
$\mancB{q}{(t_1\rss t_k)}=\{p\}$ such that ${\rt(\atpos{t_1}{p})} \not\in {\CS \cup \VS}$. 
Furthermore let $(t_i,p_i)$ be the unique node in $G$ where 
$q\in\mchildB{p_i}{(t_i\rss t_k)}$ and for any direct successor 
node $(t_j,p_j)$ of $(t_i,p_i)$ in $G$, we have $q\notin\mchildB{p_j}{(t_j\rss t_k)}$. 
Then $(t_i,p_i)$ is said to \emph{cover} the position $q \in B_k$.
\end{defi}

As shown in the next lemma, $C$ is an upper bound on the number of
positions in $B_n$ each node in $G$ can cover.
\begin{lem}
\label{lemagf:leafbranching}
For every node $(t_i,p)$ in $G$, there are at most $C$ many positions 
$q\in B_n$ covered by $(t_i,p)$.
\end{lem}
\proof
If there is no $i\leqslant j<n$ such that the redex position of the step
$t_j \rew t_{j+1}$ and an element of $\mchildB{p}{(t_i\rss t_j)}$ coincide,
then it follows from Lemma~\ref{lem:hideRsteps} that 
$\card{\mchildB{p}{(t_i\rss t_n)}} \leqslant 1 < C$.

Otherwise, let $k$ be the smallest number such that $k\geqslant i$ and
$\mchildB{p}{(t_i\rss t_k)}=\{p_k\}$, where $p_k$ is the redex position of
$t_k\rew t_{k+1}$. By Definitions \ref{def:ancestorstep} and
\ref{def:mainancestor}, $\card{\mchildB{p_k}{(t_k\rew t_{k+1})}}\leqslant C$.
In the next paragraph, we show for each $p_{k+1}\in\mchildB{p_k}{(t_k\rew t_{k+1})}$ that
$\card{\mchildB{p_{k+1}}{(t_{k+1}\rss t_n)}}\leqslant 1$. Hence the
node $(t_k,p_k)$ can cover at most $C$ many positions in $B_n$.

For each $p_{k+1}\in\mchildB{p_k}{(t_k\rew t_{k+1})}$, if
$\rt(\atpos{t_{k+1}}{p_{k+1}})$ is defined, then $(t_{k+1},p_{k+1})$ is a successor node
of $(t_i,p)$, and for any main progeny $q$ of $p_{k+1}$, by definition we have
$q\in\mchildB{p_{k+1}}{(t_{k+1}\rss t_n)}$, which violates the
definition of being covered by $(t_i,p)$.
On the other hand, suppose $\rt(\atpos{t_{k+1}}{p_{k+1}})$ is a constructor
symbol or a variable. Then by Lemma~\ref{lem:noconstructorbranching},
$\card{\mchildB{p_{k+1}}{(t_{k+1}\rss t_n)}}\leqslant1$. 
\qed

The following example illustrates the role of Lemma~\ref{lemagf:leafbranching}.
\begin{exa}
\label{ex:agraphlostdepth}
Let $\RSe$ be the TRS consisting of the single rewrite rule
\begin{equation*}
\md(\ms(x))\rew\ms(\ms(\md(x))) \tpkt
\end{equation*}
Let $t_1=\md(\ms(\ms(\Null)))$, $t_2=\ms(\ms(\md(\ms(\Null))))$, and
$t_3=\ms(\ms(\ms(\ms(\md(\Null)))))$. We have the derivation
$A\colon t_1\rew t_2\rew t_3$ and the following progenitor graph:
\begin{center}
\begin{tikzpicture}
\node (1te) {};
\node[right of=1te, node distance=3cm] (2t11) {};
\node[right of=2t11, node distance=3cm] (3t1111) {};

\draw[->] (1te) -> (2t11);
\draw[->] (2t11) -> (3t1111);

\draw (1te) node[above] (1tecaption) {$(t_1,\epsilon)$};
\draw (2t11) node[above] (2t11caption) {$(t_2,11)$};
\draw (3t1111) node[above] (3t1111caption) {$(t_3,1111)$};
\end{tikzpicture}
\end{center}
Note that $G$ leaves out all function symbols $\ms$ above
the $\md$ in each term. However, by Lemma~\ref{lemagf:leafbranching}, the number
of positions in the last term of $A$ which are hidden in this way is bounded
linearly in the size of the progenitor graph. 
\end{exa}

The next lemma shows that the ``branching factor'' of $G$, i.e., the maximal
number of direct successors of a node in $G$, is bounded by the
branching constant $C$.

\begin{lem}
\label{lemagf:nodebranching}
Every node in $G$ has at most $C$ many direct successor nodes.
\end{lem}
\proof
Let $(t_i,p)$ be a node in $G$. 
If there is no $i\leqslant j<n$ such that the redex
position of the step $t_j\rew t_{j+1}$ 
and an element of $\mchildB{p}{(t_i\rss t_j)}$
coincide, then $(t_i,p)$ has no successor node, so the claim holds.
Otherwise, let $j$ be the smallest number greater than $i$ such that
$\mchildB{p}{(t_i\rss t_j)}=\{q\}$, where $q$ is the redex position of
$t_j\rew t_{j+1}$. By Definitions \ref{def:ancestorstep} and
\ref{def:mainancestor}, $\card{\mchildB{q}{(t_j\rew t_{j+1})}}\leqslant C$.
Hence, $(t_i,p)$ has at most $C$ many direct successor nodes.
\qed

We are ready to prove the main lemma of this section.
\begin{lem}
\label{lem:dp2height}
Let $\RS$ be terminating and let $f(t) \defsym 
\max\{\dheight(u^\sharp,\rsrew{\DP(\RS)/\RS})\mid {u} \subterm {t}\}$. 
Then there exists $d \in \N$ such that for all terms $t$:
${\pdp(t,\rsrew{\RS})} \leqslant {(\depth{t}+1)\cdot2^{d \cdot (f(t) + 2)}}$.
\end{lem}
\proof
Consider any derivation $A\colon s\rssrew{\RS} t$ and
let $A' \colon (u)^\sharp \rssrew{\DP(\RS)/\RS} (v)^\sharp$ be
a maximal derivation over $\DP(\RS)$ modulo $\RS$ such that $u\subterm s$.
Set $m \defsym \dpsize{A'}$.
Let $k$ be the number of defined symbols in the main branch of $s$. If $k=0$,
then $s$ is a normal form, hence $s=t$, and the lemma follows trivially.
In the following we assume $k>0$. Note that ${k} \leqslant {\depth{s}+1}$. 
It is easy to see that the progenitor graph $G$ 
forms a forest consisting of $k$ distinct trees $T_1,\dots,T_k$. 

Due to Lemma~\ref{lemagf:edgederivation} 
the height of each tree $T_1,\dots,T_k$ in $G$ is bounded by $m$. 
Here the height of a tree is the number of edges on the 
longest path from the root to a leaf. 
Recall that any $C$-ary tree of height $m$
has at most $\frac{C^{m+1}-1}{C-1} \leqslant C^{m+1}$ many nodes. 
Hence, due to Lemma~\ref{lemagf:nodebranching}, each of the
trees $T_i$ ($1 \leqslant i \leqslant k$) has at most $C^{m+1}$ many
nodes. Thus $G$ can have at most $k \cdot C^{m+1}$ many nodes.

The main branch of $t$ consists of $\depth{t}+1$ many positions,
all of which have to fulfil one of the two properties in 
Lemma~\ref{lemagf:allfinalcovered}.
By Lemma~\ref{lem:noconstructorbranching}, the first case applies
to at most $\depth{s}+1-k$ many positions. Furthermore, due
to Lemma~\ref{lemagf:leafbranching} each node in $G$ can cover
at most $C$ many positions in the main branch of $t$. 
In sum we obtain the following upper bound on the depth of $t$:
\begin{equation*}
\depth{t} \leqslant (k \cdot C^{m+1}) \cdot C + \depth{s} - k
\leqslant (\depth{s}+1)\cdot C^{m+2} \tkom
\end{equation*}
where we have applied $k \leqslant \depth{s} +1$ in the
second inequality.
By definition $m = \dpsize{A'} = f(s)$ 
from which the lemma follows immediately.
\qed

All that is left to show is that the derivational complexity of a finite and
terminating TRS is bounded double exponentially in its depth growth. This
can be achieved by two easy observations.
\begin{lem}
\label{lem:height2size} \label{lem:size2dc}
Let $\RS$ be terminating. Then there exists $d \in \N$ such that
for all terms $t$:
$\dheight(t,\rsrew{\RS})\leqslant 2^{2^{d\cdot\pdp(t,\rsrew{\RS})}}$.
\end{lem}
\proof
We show that there exist constants $e$ and $e'$, such that for all
terms $t$, the inequalities $\psz(t,\rsrew{\RS})\leqslant 2^{e\cdot\pdp(t,\rsrew{\RS})}$ and
$\dheight(t,\rsrew{\RS})\leqslant2^{e'\cdot\psz(t,\rsrew{\RS})}$ hold. Then the lemma
follows easily by choosing $d=e+e'$, for instance.
\begin{enumerate}[(1)]
\item For any term $t$, we have $\size{t}\leqslant k^{\depth{t}+1}$, where $k$
is the maximum arity of any function symbol in the signature. This proves the
first inequality.
\item On the other hand, by assumption the signature $\FS$ of $\RS$ is finite.
Moreover without loss of generality the considered derivation in $\RS$ is ground.
Hence we can build only $2^{e'\cdot m}$ different terms of size at most $m$,
where $e'$ depends only on $\FS$. This proves the second inequality.
\qed
\end{enumerate}
\begin{thm}
\label{thm:elemwall}
Let $\RS$ be terminating and let 
\begin{equation*}
 f(n) \defsym \max\{ \dheight(t^\sharp,\rsrew{\DP(\RS)/\RS}) \mid \size{t} \leqslant n\} \tpkt 
\end{equation*}
Then there exists $D \in \N$ such that $\Dc{\RS}(n)\leqslant 2^{2^{n\cdot2^{D\cdot(f(n)+2)}}}$.
\end{thm}
\proof
The theorem follows directly from Lemmata~\ref{lem:dp2height} and \ref{lem:size2dc}.
\qed

We also call the function $f$ defined in the theorem the \emph{dependency pair complexity}
function. Observe that for any argument filtering $\pi$ and any terms $s$, $t$, we have
that  $s^\sharp\rsrew{\RS}t^\sharp$ implies $\pi(s^\sharp)\rsgrew{\pi(\RS)}\pi(t^\sharp)$.
Furthermore $s^\sharp\rsrew{\DP(\RS)}t^\sharp$ implies
$\pi(s^\sharp)\rsrew{\pi(\DP(\RS))}\pi(t^\sharp)$. These observations are sufficient
to extend Theorem~\ref{thm:elemwall} to argument filtering.

\begin{cor}
\label{t:af}
Let $\RS$ be terminating, let $\pi$ be an argument filtering, and let 
\begin{equation*}
f(n) \defsym \max\{ \dheight(\pi(t^\sharp),\rsrew{\pi(\DP(\RS))/\pi(\RS)}) \mid \size{t} \leqslant n\} \tpkt  
\end{equation*}
Then $\Dc{\RS}(n)\leqslant 2^{2^{n\cdot2^{D\cdot(f(n)+2)}}}$, where
$D$ is defined as above.
\qed
\end{cor}

This concludes that termination proofs by the basic dependency pair method
combined with some base technique (possibly enhanced by argument filtering)
imply a complexity bound that is triple exponential in the derivational
complexity of the base technique. For instance, if polynomial interpretations
are used as base technique, then the derivational complexity of the TRS under
consideration is bounded by $2_5(\bigO(n))$. On the other hand, if KBO is used
as a base technique, then the derivational complexity of the TRS is bounded by
$\Ack(\bigO(n),0)$.

So by Theorem~\ref{thm:elemwall}, the derivational complexity of a TRS $\RS$
is bounded triple exponentially in its dependency pair complexity. This yields
an upper-bound. The following TRS establishes a double exponential lower-bound.

\begin{exa}
\label{ex:dplowerbound}
Consider the following TRS $\RSde$, extending the TRS $\RSd$:
\begin{alignat*}{4}
1\colon &\;& \mfull(\ms(x)) & \rew\ms(\mfull(\mfull(x))) & \hspace{10ex}
2\colon &\;& \mfull(x) &\rew\mcs(x,x) \tpkt
\end{alignat*}
We show that $\RSde$ has linear dependency pair complexity, but
admits derivations of double exponential length.

Let $C(x)$ be the shorthand for $\mcs(x,x)$.
Now, consider the starting term $\mfull(\ms^n(\Null))$. As can be easily seen, this
term rewrites to $\ms^n(\mfull^{2^n}(\Null))$ in $2^n-1$ steps using rule~1. Now,
we can use rule~2 and an outermost strategy to reach $\ms^n(C^{2^n}(\Null))$
in $2^{2^n}-1$ steps, so $\Dc{\RSde}$ is at least double exponential.

On the other hand consider $\DP(\RSde)$:
\begin{alignat*}{4}
3\colon &\;& \mfull^\sharp(\ms(x)) &\rew \mfull^\sharp(\mfull(x)) & \hspace{10ex}
4\colon &\;& \mfull^\sharp(\ms(x)) &\rew\mfull^\sharp(x) \tpkt
\end{alignat*}
We define a (very restricted) polynomial interpretation $\AA$ as follows:
$\mfull^\sharp_\AA(m) = \mfull_\AA(m) = m$, $\ms_\AA(m)= m + 1$, $\mcs_\AA(m,n) = \Null_\AA = 0$,
where ${\RSde} \subseteq {\succcurlyeq_\AA}$ and
${\DP(\RSde)} \subseteq {>_\AA}$, and $(\succcurlyeq_\AA,>_\AA)$ forms a reduction pair.
Thus the dependency pair complexity function with respect to $\RSde$ is at most linear.
\end{exa}

Note that from the proof of Theorem~\ref{thm:elemwall} one can distill the following
three facts, where each of them is responsible for one of the 
exponentials in the upper-bound:
\begin{enumerate}[(1)]
\item \label{en:lb:1}
the number of nodes in a progenitor graph may be exponential in its height,
\item \label{en:lb:2}
the size of a term may be exponential in its depth, and
\item \label{en:lb:3}
the number of terms of size $n$ is exponential in $n$.
\end{enumerate}
For an optimal example, we would have to utilise
all three criteria, while the just given TRS $\RSde$ utilises only the
criteria~(\ref{en:lb:1}) and (\ref{en:lb:2}).
To us, it seems impossible to enumerate enough
terms of exponential depth and double exponential size 
so that this is possible. Moreover, we believe that the first and
the last criterion can be merged into a single exponential, as shown
for string rewriting in the next section. Hence, we conjecture that
the upper-bound given in Theorem~\ref{thm:elemwall} can be improved
to a double exponential one.

\section{String Rewriting}
\label{sec:srscomplexity}

In this short section we consider \emph{string rewrite systems} 
(\emph{SRSs} for short), i.e., TRSs where all function symbols are unary or nullary.%
\footnote{This is sometimes called \emph{unary rewriting}, as opposed
to ``true'' string rewriting, where only unary function symbols are
allowed. The results presented in this section hold for both flavours
of string rewriting.}
Since the size and the height of strings are
linearly (and not just exponentially) related, the upper bound from
Theorem~\ref{thm:elemwall} immediately breaks down to a double exponential
one. However, we can further improve this bound to a single exponential
one. Showing this is the purpose of this section. Our main proof
idea is to relate rewrite steps and nodes in the progenitor graph directly.

As in Section~\ref{sec:dpcomplexity}, we fix a finite SRS $\RSS$, a
derivation $A:t_1\rss t_n$ over $\RSS$, and the progenitor graph $G$
of $A$. In this section, the terms $t_1,\ldots,t_n$ do not contain any
function symbols with arity greater than $1$. Therefore, each of
them only consists of a single branch, which in turn is its main
branch.
\begin{lem}
\label{lem:step2node}
The number of nodes in $G$ is at least $n-1$.
\end{lem}
\proof
For each $1\leqslant k\leqslant n-1$, let $p_k$ be the redex
position of the step $t_k\rew t_{k+1}$, so that $\rt(\atpos{t_k}{p_k})$
is defined.

Since we consider string rewriting, $p_k$ is in the main branch
of $t_k$. Hence by Lemma~\ref{lemagf:allfinalcovered}, there exists
a node $(t_i,p)$ in $G$ that covers $(t_k,p_k)$. Moreover, for any
$j$ ($i \leqslant j < k$), the single element of
$\mancB{p_k}{(t_j\rss t_k)}$ and $p_j$ do not coincide. Otherwise
$(t_j,p_j)$ would be a successor of $(t_i,p)$ that
covers $(t_k,p_k)$. This would contradict the choice
of $(t_i,p)$. 
Therefore, by Lemma~\ref{lem:hideRsteps}, we have
$\mchildB{p}{(t_i\rss t_k)}=\{p_k\}$. This yields a one to one
correspondence between all $n-1$ redex positions and a subset
of the nodes of $G$, entailing the lemma.
\qed

\begin{thm}
\label{thm:srsdp2dc}
Let $\RSS$ be terminating and let
$f(n) \defsym \max\{\dheight(t^\sharp,\rsrew{\DP(\RSS)/\RSS})\mid \size{t} \leqslant n\}$.
Then there exists $d \in \N$ such that
$\Dc{\RSS}(n) \leqslant n\cdot2^{d \cdot (f(n)+1)}$.
\end{thm}
\proof
Recall the branching constant $C = \max(\{2\}\cup\{\depth{r}+1\mid l\rew r\in\RSS\})$.
Let $A \colon s\rew^n_{\RSS} t$ denote any derivation with respect to
$\RSS$.

Let $G$ be the progenitor graph of $A$. 
$G$ has $k$ many connected components, where $k$ is the number of defined
symbols in $s$. By Lemma~\ref{lemagf:nodebranching},
each of them contains at most $C^{\dpsize{A'}+1}$ many nodes, where $A'$ is an
implicit dependency pair derivation corresponding to $A$ such that $\dpsize{A'}$
is maximal.
Hence the total size of $G$ is most $k\cdot C^{\dpsize{A'}+1}$. By
Lemma~\ref{lem:step2node}, the size of $G$ is at least $n-1$, 
from which we obtain:
\begin{equation*}
 n\leqslant k\cdot C^{\dpsize{A'}+1} + 1\leqslant \size{s}\cdot C^{\dpsize{A'}+1} + 1\tpkt
\end{equation*}
We obtain that the length $n$ of $A$ is less than or equal to
$\size{s} \cdot C^{\dpsize{A'}+1} + 1$. From this the theorem is immediate.
\qed

Thus, for string rewriting, termination proofs by the basic dependency pair method
combined with some base technique and an argument filtering
imply a complexity bound that is only single exponential in the derivational 
complexity of the base technique. If polynomial interpretations
are used as base technique, then the derivational complexity of the SRS under
consideration is bounded by $2_3(\bigO(n))$. If the base technique is KBO,
then the derivational complexity of the SRS is bounded by $\Ack(\bigO(n),0)$.

\begin{exa}
\label{ex:srslowerbound}
If we restrict $\RSde$ to its first rule (i.e., we consider $\RSd$), we can
see in the same way as in Example~\ref{ex:dplowerbound} that $\Dc{\RSd}$ is
at least exponential, and the dependency pair complexity function with respect
to $\RSd$ is at most linear. Therefore, the upper bound given in Theorem~\ref{thm:srsdp2dc} is tight.
\end{exa}

This concludes our complexity analysis of the basic
dependency pair method. The purpose of the next sections is to analyse the
usable rules and dependency graph refinements.

\section{Usable Rules}
\label{sec:ur}

In this section we extend the results in Section~\ref{sec:dpcomplexity} to
termination proofs employing the dependency pair method in connection with the
usable rules criterion, c.f.~Proposition~\ref{prop:ur}.

\begin{notation}
For the rest of this section, we use the following constants depending only
on the TRS $\RS$. 
Let $E \defsym \max\{a,b,2 \cdot c\}+3$, where $a$ is the maximum arity
of all function symbols, $b$ the number of rules in $\RS$, $c$ is chosen
such that it is larger than the size of any right hand side of any rule in $\RS$, and
hence also larger than the number of occurrences of any variable on the right
hand side. Furthermore let $F \defsym \max \{a,D\}$,
where $D$ is defined as in Theorem~\ref{thm:elemwall}.
\end{notation}

Perhaps surprisingly, the usable rules criterion
strengthens the power of the termination technique considerably, as witnessed
by the following example.

\begin{exa}
\label{ex:ur}
Consider the TRS $\RSebin$:
\begin{alignat*}{4}
&&\md(\Null)&\rew\Null & 
&&\me(\Null,x)&\rew x
\\
&&\md(\ms(x))&\rew\ms(\ms(\md(x))) & \hspace{8ex}
&&\me(\ms(x),y)&\rew\me(x,\md(y)) 
\tpkt
\end{alignat*}
The dependency pairs of $\RSebin$ are given by
$\md^\sharp(\ms(x))\rew\md^\sharp(x)$,
$\me^\sharp(\ms(x),y)\rew\me^\sharp(x,\md(y))$, and
$\me^\sharp(\ms(x),y)\rew\md^\sharp(y)$.

We have the following usable rules with respect to $\RSebin$:
\begin{equation*}
  \md(\Null)\rew\Null \qquad \md(\ms(x))\rew\ms(\ms(\md(x))) \tpkt
\end{equation*}
The rules $\DP(\RSebin)\cup\U(\DP(\RSebin))\cup \Ce$
admit only double exponentially
many dependency pair steps from any starting term
$t^\sharp$ with $t\in\TERMS$. Consider for instance the algebra $\A$ over $\N$
defined as follows:
\begin{alignat*}{3}
\me^\sharp_\A(m,n) & =2^{m}\cdot(n+1)+1 & \hspace{5ex}
\md^\sharp_\A(m) & =m & \hspace{5ex}
\md_\A(m) & = 2\cdot m
\\
\Null_\A & = 0 & \hspace{5ex}
\ms_\A(m) & = m+1 & \hspace{5ex}
\mcs_\A(m,n) & = m+n
\tpkt
\end{alignat*}

\noindent It is easy to check that $\DP(\RSebin) \subseteq {>_\A}$ and
$\U(\DP(\RSebin))\cup\Ce \subseteq {\geqslant_\A}$, and for any term $t\in\TERMS$,
$\eval{\alpha}{t^\sharp}$ is double exponentially bounded in $\size{t}$.

On the other hand the derivational complexity with respect to $\RSebin$ is
clearly super-exponential. Observe that also the rules 
$\DP(\RSebin)\cup\RSebin$ allow a super-exponential number
of $\DP$-steps, e.g. for the family of starting terms $\me^\sharp(E^k(\Null),\ms(\Null))$,
where $E(x)$ is a shorthand for $\me(x,\ms(\Null))$.
\end{exa}

Note that in Example~\ref{ex:ur} it is essential
that arbitrary starting terms, as for example 
$\me(E^k(\Null),\ms(\Null))$, are considered. 
If we would restrict 
the starting terms to \emph{basic} terms, i.e., terms of the form $f(\seq{t})$ 
such that $f$ is defined and $t_i\in\VALUES$ 
for all $1 \leqslant i \leqslant n$, then 
the results from Section~\ref{sec:dpcomplexity} directly extend to
Proposition~\ref{prop:ur}. This is a consequence of \cite[Lemma 16]{HM08a}.
We can generalise Example~\ref{ex:ur} to primitive recursion by employing the
Ackermann function. 

\begin{exa}
\label{ex:urgen}
We employ a unary notation for the Ackermann
function: we write $\Ack_i(x)$ instead of $\Ack(i,x)$. Consider the
following family of TRSs $\RSaack(l)$, parametrised by $l\in\N$.
Here we assume $0 \leqslant i < l$.
\begin{alignat*}{4}
&& \mAck_0(x) &\rew \ms(x) & 
&& \mAAck(\Null,x) &\rew \mAck_l(x) 
\\
&&\mAck_{i+1}(\Null) &\rew \mAck_i(\ms(\Null)) & 
&& \mAAck(\ms(x),y) &\rew \mAAck(x,\mAck_l(y))
\\
&&\mAck_{i+1}(\ms(x)) &\rew \mAck_i(\mAck_{i+1}(x)) \tpkt & \hspace{8ex} 
\end{alignat*}

Note that the last two rules are not contained in $\U(\DP(\RSaack(l)))$. The
number of dependency pair steps admitted by the rules $\DP(\RSaack(l))\cup\U(\DP(\RSaack(l)))\cup\Ce$
from any starting term $t^\sharp$ with $t\in\TERMS$
is then bounded by $\Ack^2_{l+1}(\bigO(\size{t}))$, as witnessed by the algebra
$\A$ over $\N$, defined as follows. 
\begin{alignat*}{3}
(\mAck_i^\sharp)_\A(m) &=\Ack_i(m)+i & \hspace{3ex}
\mAAck^\sharp_\A(m,n) & =\Ack_l^{m+1}(n)+m+l+1 & \hspace{3ex}
\ms_\A(m) & =m+1 
\\
(\mAck_i)_\A(m) & =\Ack_i(m) & 
\mcs_\A(m,n) &=m+n &
\Null_\A & =0 
\end{alignat*}

\noindent It is easy to check that $\DP(\RSaack(l)) \subseteq {>_\A}$ and
$\U(\DP(\RSaack(l)))\cup\Ce \subseteq {\geqslant_\A}$.
On the other hand the derivational complexity with respect to $\RSaack(l)$ is
bounded from below by $\Ack_{l+2}(\Omega(n))$, as witnessed by derivations
starting from the family of terms $F^k(\Null)$, where $F(x)$ is a
shorthand for $\mAAck(x,\ms(\Null))$. 

It follows that the derivational complexity of the base technique used in the
termination proof of $\bigcup_{i=0}^{l} \RSaack(l)$ belongs to level
$l+1$ of the Ackermann function, while the derivational complexity of the considered TRS
belongs to level $l+2$.
\end{exa}

Due to Examples~\ref{ex:ur} and \ref{ex:urgen} 
we cannot have an elementary relationship between
the derivational complexity of the original TRS $\RS$ and the complexity induced
by the termination technique employed in conjunction with Proposition~\ref{prop:ur}.

Still, we can give an upper bound on the derivational complexity with respect to $\RS$. 
This follows from a close study of the correctness proof of Proposition~\ref{prop:ur}
given in~\cite{HM07}, compare also \cite{GTSF06}.
The main ingredient of this proof is the definition of the interpretation
$\IG$.
\begin{defi}[\cite{HM07}]
\label{def:ig}
Let $\GS\subseteq\FS$. The
\emph{interpretation} $\IG$ is a mapping from terminating terms in
$\TA(\FS^\sharp,\VS)$ to terms in $\TA(\FS^\sharp\cup\{\mnil,\mcs\},\VS)$,
where $\mnil$ is a fresh function symbol and $\mcs$ is the function symbol
introduced by $\Ce$, inductively defined as follows:
\begin{equation*}
\IG(t)\defsym\left\{\begin{array}{ll}
t & \text{if $t$ is a variable}\\
f(\IG(t_1),\ldots,\IG(t_n)) & \text{if $t=f(t_1,\ldots,t_n)$ and $f\notin\GS$}\\
\mcs(f(\IG(t_1),\ldots,\IG(t_n)),t') & \text{if $t=f(t_1,\ldots,t_n)$ and $f\in\GS$}
\end{array}\right.
\end{equation*}
where in the last clause $t'$ denotes the term $\morder(\{\IG(u)\mid t\rsrew{\RS}u\})$ with
\begin{equation*}
\morder(T)\defsym\left\{\begin{array}{ll}
\mnil & \text{if $T=\varnothing$}\\
\mcs(t,\morder(T - \{t\})) & \text{if $t$ is the minimum element of $T$}
\end{array}\right.
\end{equation*}
Here an arbitrary but fixed total order on 
$\TA(\FS^\sharp\cup\{\mnil,\mcs\},\VS)$ is assumed.
\end{defi}

According to \cite[Theorem 20]{HM07}, any $\DP(\RS)\cup\RS$-derivation starting from $t$ can be transformed
into a $\DP(\RS)\cup\U(\DP(\RS))\cup\Ce$-derivation starting from $\IG(t)$,
where $\GS$ is the set of defined symbols of $\RS - \U(\DP(\RS))$.
Therefore, estimating $\size{\IG(t)}$ is the key to the connection between
$\dheight(t,\rsrew{\DP(\RS)/\RS})$ and $\dheight(t,\rsrew{\DP(\RS)/\U(\DP(\RS))\cup\Ce})$. Suppose there exists a function $f$ that bounds 
$\dheight(t,\rsrew{\DP(\RS)/\U(\DP(\RS))\cup\Ce})$ in $\size{t}$. 
Then $\dheight(t,\rsrew{\DP(\RS)/\RS})$ can be bounded in $\size{t}$ by
$f(\size{\IG(t)})$. 

However, the difficulty of this estimation lies in the
following mutual dependence between the definition of the interpretation $\IG$
and the derivation height. On one hand, we bound
$\dheight(t,\rsrew{\DP(\RS)/\RS})$ in $\size{t}$ by $f(\size{\IG(t)})$. On the
other hand, $\IG(t)$ depends on $\dheight(t,\rsrew{\RS})$ since
$\dheight(t,\rsrew{\RS})$ determines the number of recursive calls of the shape
$\{\IG(u)\mid t\rsrew{\RS}u\}$ in the definition of $\IG(t)$. 
The following sequence of lemmata shows how this mutual dependence can
be resolved. 

\begin{defi}
Let $g \colon \N\times\N\to\N$ be the function satisfying
the following recursive definition:
\begin{equation*}
g(m,n)\defsym
\begin{cases}
E & \text{if $m=0$}\\
E\cdot g(m-1,0) & \text{if $m>0$ and $n=0$}\\
E\cdot g(m-1,n) + E\cdot m \cdot g(E\cdot m,n-1) & \text{otherwise}
\tpkt  
\end{cases}
\end{equation*}  
\end{defi}\medskip

\noindent The next two lemmata estimate the size $\size{\IG(t)}$ of the
interpretation $\IG(t)$ in the size of $t$ and the derivation height of $t$
(with respect to $\RS$).
\begin{lem}
\label{lem:boundig}
\mbox{}
\begin{enumerate}[\em(1)]
\item The function $g$ is well-defined and strictly monotone in each
argument.
\item For all $m$, $n$: $E \leqslant g(m,n)$.
\item \label{en:lem:boundig}
For any term $t$: $\size{\IG(t)}\leqslant g(\size{t},\dheight(t,\rsrew{\RS}))$.
\end{enumerate}
\end{lem}
\proof
We only show property~(\ref{en:lem:boundig}). The proof 
proceeds by induction on the lexicographic order over the
pair $(\dheight(t,\rsrew{\RS}),\size{t})$. 
It suffices to consider the
interesting case, where $t = f(t_1,\ldots,t_n)$ with $f\in\GS$
and $\dheight(t,\rsrew{\RS})>0$.
We obtain 
\begin{align*}
\size{\IG(t)}
&=2+\sum_{i=1}^n\size{\IG(t_i)}+\size{\morder(\{\IG(u)\mid t\rsrew{\RS}u\})}\\
&\leqslant 2+\sum_{i=1}^n\size{\IG(t_i)}+1
+b\cdot \size{t} \cdot(1+\max\{\size{\IG(u)}\mid t\rsrew{\RS}u\})\\
&\leqslant 3+
n\cdot \max_{1 \leqslant i \leqslant n} \{g(\size{t_i},\dheight(t_i,\rsrew{\RS}))\}
+b\cdot \size{t} \cdot(1+\max\{g(\size{u},\dheight(u,\rsrew{\RS})) \mid t\rsrew{\RS}u\})\\
& \leqslant 3+ a \cdot g(\size{t}-1,\dheight(t,\rsrew{\RS}))
+b\cdot \size{t}
+b\cdot\size{t} \cdot g(c\cdot\size{t}+c,\dheight(t,\rsrew{\RS})-1)\\
&\leqslant E\cdot g(\size{t}-1,\dheight(t,\rsrew{\RS}))
+E\cdot\size{t}\cdot g(E\cdot\size{t},\dheight(t,\rsrew{\RS})-1)\\
&=g(\size{t},\dheight(t,\rsrew{\RS}))
\tpkt
\end{align*}
In the second line we use the fact that any term $t$ has at most $b\cdot\size{t}$
many reducts. In the third line, we apply the induction
hypothesis. In the fourth line, we use that $\size{u}\leqslant c\cdot\size{t}+c$
whenever $t\rsrew{\RS}u$.
\qed
\begin{lem}
\label{lem:helem}
Let $g$ be defined as in Lemma~\ref{lem:boundig} above. Then there exists
a minimal $d \in \N$ such that for all $m,n \in \N$, we have
$g(m,n)\leqslant 2^{2^{d\cdot(m+n+1)}} =: G(m,n)$.
\end{lem}
\proof
It can be shown by straightforward induction on the lexicographic order over
the pair $(m,n)$ that $g(m,n)\leqslant (E\cdot(n+1))^{(n+1)\cdot E^{2\cdot m+1}}$
holds.
It is easy to see that for suitable $d$ we have
$(E\cdot(n+1))^{(n+1)\cdot E^{2\cdot m+1}}\leqslant 2^{2^{d\cdot(m+n+1)}}$. 
Thus the lemma follows.
\qed

For the remainder of the section, let the function $G$ be defined as in
Lemma~\ref{lem:helem} above. Then we define the 
function $H[f] \colon \N \to \N$ parametrised in a mapping $f$ from the
naturals to the naturals as follows:
\begin{equation*}
H[f](m) \defsym f(1+F\cdot G(m,h(m,m))) \tkom
\end{equation*}
where $h(m,n)\defsym2^{2^{m\cdot2^{F\cdot (n+2)}}}$.

\begin{lem}
\label{lem:urimpact}
Let $\RS$ be terminating and let the function $f$ be defined by
$f(n) \defsym \max (\{n\} \cup \{\dheight(t^\sharp,\rsrew{\DP(\RS)/\U(\DP(\RS)) \cup \Ce}) \mid \size{t} \leqslant n\})$.
Then $\dheight(t^\sharp,\rsrew{\DP(\RS)/\RS}) \leqslant (H[f])^{\size{t}}(1)$.
\end{lem}
\proof
We show the lemma by induction on $t^\sharp$. If $t^\sharp = t$ is a variable, the lemma
is trivial. Otherwise, assume $t^\sharp = f^\sharp(t_1,\ldots,t_n)$ and
recall that $f^\sharp\notin\GS$ for any $f\in\FS$.
Due to Definition~\ref{def:ig} in conjunction with Lemma~\ref{lem:boundig}
there exists $i \in \{1,\dots,n\}$ such that
\begin{equation*}
\size{\IG(t^\sharp)} = \size{f^\sharp(\IG(t_1),\dots,\IG(t_n))}
\leqslant 1+F\cdot g(\size{t_i},\dheight(t_i,\rsrew{\RS})) \tpkt
\end{equation*}
By Theorem~\ref{thm:elemwall}, we have
\begin{equation*}
\dheight(t_i,\rsrew{\RS}) \leqslant
2^{2^{\size{t_i}\cdot 2^{D\cdot(\dheight(t^\sharp_i,\rsrew{\DP(\RS)/\RS})+2)}}}
\tpkt
\end{equation*}
Due to $D \leqslant F$, we conclude
$\dheight(t_i,\rsrew{\RS}) \leqslant h(\size{t_i},\dheight(t^\sharp_i,\rsrew{\DP(\RS)/\RS}))$.

As mentioned above, any $\DP(\RS)\cup\RS$-derivation starting from $t$ can be transformed
into a $\DP(\RS)\cup\U(\DP(\RS))\cup\Ce$-derivation starting from $\IG(t)$.
Thus, we also have $\dheight(t^\sharp,\rsrew{\DP(\RS)/\RS})\leqslant f(\size{\IG(t^\sharp)})$.
In sum we obtain:
\begin{align*}
\dheight(t^\sharp,\rsrew{\DP(\RS)/\RS})
&\leqslant f(\size{\IG(t^\sharp)})\\
&\leqslant f(1+F\cdot g(\size{t_i},\dheight(t_i,\rsrew{\RS})))\\
&\leqslant f(1+F\cdot g(\size{t_i},h(\size{t_i},\dheight(t_i^\sharp,\rsrew{\DP(\RS)/\RS}))))\\
&\leqslant f(1+F\cdot G(\size{t_i},h(\size{t_i},\dheight(t_i^\sharp,\rsrew{\DP(\RS)/\RS}))))\\
&\leqslant H[f](\max\{\size{t_i},\dheight(t_i^\sharp,\rsrew{\DP(\RS)/\RS})\}) \tpkt
\end{align*}
It is easy to verify that $\size{t_i}\leqslant (H[f])^{\size{t_i}}(1)\leqslant (H[f])^{\size{t}-1}(1)$.
Moreover by induction hypothesis we have
$\dheight(t_i^\sharp,\rsrew{\DP(\RS)/\RS}) 
\leqslant (H[f])^{\size{t_i}}(1) 
\leqslant (H[f])^{\size{t}-1}(1)$. From this the lemma
follows.
\qed

\begin{thm}
\label{thm:urimpact}
Let $\RS$ be terminating and let 
\begin{equation*}
f(n) \defsym \max (\{n\} \cup \{\dheight(t^\sharp,\rsrew{\DP(\RS)/\U(\DP(\RS)) \cup \Ce}) \mid \size{t} \leqslant n\}) \tpkt
\end{equation*}
Then there exist a function $f'$ which is elementary in $f$, and an elementary
function $e$ such that $\Dc{\RS}(n) \leqslant e(n,(f')^{n}(1))$.
\end{thm}
\proof
We choose $f'=H[f]$ and $e=h$.
Let $t$ be a term. By Theorem~\ref{thm:elemwall} we have that 
$\dheight(t,\rsrew{\RS}) \leqslant h(\size{t},\dheight(t^\sharp,\rsrew{\DP(\RS)/\RS}))$.
Furthermore by Lemma~\ref{lem:urimpact} we obtain that 
$\dheight(t^\sharp,\rsrew{\DP(\RS)/\RS}) \leqslant (H[f])^{\size{t}}(1)$. Combining
these two observations the theorem is immediate.
\qed

Consider any TRS $\RS$ whose termination can be shown by the basic dependency
pair method and some base technique enhanced by the usable rules criterion.
Let $f$, $e$, and $f'$ be defined as in Theorem~\ref{thm:urimpact}
and set $j(n) \defsym e(n,(f')^{n}(1))$.
For instance, if polynomial interpretations are used as a base technique,
then $f$ is bounded by a double exponential function. Therefore $j$ (and thus
also $\Dc{\RS}$) is superexponentially bounded.
On the other hand, if LPO is used as a base technique, then $f$, and hence
also $j$ and $\Dc{\RS}$ are bounded by multiply recursive functions.
Note that the derivational complexity induced by LPO (as a direct method) is multiply recursive~\cite{W95}.
Clearly the class of  multiply recursive functions is closed under
primitive recursion. Hence the complexity of the dependency pair method (in conjunction
with the usable rules refinement) becomes negligible.

\section{Dependency Graphs}
\label{sec:dg}

We now consider \emph{dependency graphs}, i.e., 
we establish an upper bound on the complexity of TRSs whose termination can be shown by Proposition~\ref{prop:dg}. As already mentioned in the introduction 
the derivational complexity analysis of Proposition~\ref{prop:dg} does not 
employ the techniques developed in 
Sections~\ref{sec:dpcomplexity}--\ref{sec:ur}, but a conceptually
simpler technique. Essentially it suffices to
embed the TRS $\RS$ in a generic simulating TRS $\RSsim$ 
(see Definition~\ref{d:1}), whose derivational complexity can be analysed directly.

\begin{notation}
For the rest of this section, we use the following constants depending only
on the TRS $\RS$.
Let $k$ be the number of (trivial and nontrivial) SCCs in $\DG(\RS)$,
$a$ the maximum arity of any function symbol
occurring in $\RS$, and recall that $C$ denotes the branching constant of $\RS$.
\end{notation}

At first glance, it might seem that the number of dependency pair steps admitted
by a TRS is bounded linearly in the number of dependency pair steps admitted
within the ``worst'' SCC of the dependency graph. 
However, this is not the case.
\begin{exa}
\label{ex:dglowerbound}
Consider the following family of TRSs, denoted as $\RSdieterl(l)$, and parametrised
by $l \in \N$. The system of TRSs $\RSdieterl(l)$ generalises a TRS given 
by Hofbauer in~\cite{H92b}.
\begin{align*}
\mi(x)\mcirc_k (y\mcirc_{k-1}z) &\rew x \mcirc_k(\mi(\mi(y))\mcirc_{k-1}z)
& & 2 \leqslant k \leqslant l\\
\mi(x)\mcirc_k (y\mcirc_{k-1}(z\mcirc_{k-2}w)) &\rew x\mcirc_k(z\mcirc_{k-1}(y\mcirc_{k-2}w))
& & 3 \leqslant k \leqslant l
\end{align*}
For all $m,n \geqslant 0$, set
\begin{equation*}
  t_{m,n} \defsym \mi^{2(n+1)}(\me)\mcirc_{m+2}(\me\mcirc_{m+1}(\ldots(\me\mcirc_1\me)\ldots))
\tpkt
\end{equation*}
Then $\dheight(t_{m,n},\rsrew{\RSdieterl(l)}) \geqslant \Ack(m,n)$,
whenever $l \geqslant m+2$. This follows from Proposition~5.9 in~\cite{H92b}.
Hence, for every
primitive recursive function $f$ there exists some $l$ such that $\Dc{\RSdieterl(l)}$
dominates $f$. Due to Theorem~\ref{thm:elemwall} the same property
holds for the dependency pair complexities of the TRSs $\RSdieterl(l)$.

On the other hand, we can show termination of $\RSdieterl(l)$ by orienting every
nontrivial SCC of $\DG(\RSdieterl(l))$ by a uniform and restricted polynomial interpretation $\A$. We define $(\mcirc^\sharp_k)_\A(m,n) = m$,
$(\mcirc_k)_\A(m,n) = 0$,
$ \mi_\A(m) = m+1$, where $k \in\{1,\ldots,l\}$.
Note that $\A$ yields a linear upper bound on the number of dependency pair steps in each SCC.
\end{exa}

\begin{rem}
In \cite[Section 6]{MS09} we falsely claimed that the derivational complexity
induced by Proposition~\ref{prop:dg} would be \emph{elementary} in the complexity of
the base techniques. Example~\ref{ex:dglowerbound} contradicts this claim.
\end{rem}

Example~\ref{ex:dglowerbound} exemplifies the fact that the bound on the
maximal number of dependency pair steps possible within 
a specific SCC $\PP$ is related to the size of the
considered term at the moment of entering the SCC $\PP$, and not to the size of the
starting term of the full derivation. Thus, Theorem~\ref{thm:elemwall} and the
techniques developed in the preceding sections, cannot
be applied directly. Instead, one needs to argue inductively so that in each step 
in this induction, Theorem~\ref{thm:elemwall} is on one hand employed to estimate the number of
$\RS$-steps and on the other hand used to provide an upper bound on the size of
terms. 

However, this inductive argument becomes rather involved. Thus
we establish a new technique in this section, where we employ a
simulating TRS $\RSsim$. In this way the inductive argument
becomes hidden in the termination proof of $\RSsim$. 
The argument needs some preparations. Let $f$ be a monotone function over $\N$ defined as follows:
\begin{equation}
\label{eq:scccomplexity}
  f(n) \defsym \max (\{1\} \cup \{ \dheight(t^\sharp,\rsrew{\C/\RS})
  \mid \size{t} \leqslant n, \text{$\C$ is SCC of $\DG(\RS)$}\}) \tkom
\end{equation}
such that $f$ dominates the maximal number of $\PP/\RS$ steps in any SCC $\PP \in \DG(\RS)$.
Consider the unary function $f$ defined in~\eqref{eq:scccomplexity}. 
Note that it is an easy task to define a TRS $\RS'$ 
(employing the constructors $\ms$, $\Null$) that computes 
the function $f$, whenever $f$ is computable. 
That is, there exist a TRS $\RS'$ and a defined function symbol $\mf$ 
such that $\mf(\ms^n(\Null)) \rssrew{\RS'} \ms^{f(n)}(\Null)$. 

Note that if $f$ is a primitive recursive function
it is straightforward to define the TRS $\RS'$ in such a way that
the derivational complexity function $\Dc{\RS'}$ 
is primitive recursive~\cite{H92}. Furthermore it is 
not difficult to see that this generalises to any class of (computable) 
functions~\cite{HW94}.

Let $\PP$ and $\QQ$ denote different (trivial or nontrivial) SCCs in $\DG(\RS)$, respectively. Then
we call $\QQ$ \emph{reachable} from $\PP$ if there exist nodes $u \in \PP$,
$v \in \QQ$ and a path in $\DG(\RS)$ connecting $u$ with $v$.
Let $\QQ_1,\QQ_2,\dots,\QQ_k$ be all (trivial and nontrivial) SCCs in
$\DG(\RS)$. Let $\rk \colon \{\QQ_1,\dots,\QQ_k\} \to \{1,\dots,k\}$ be
an arbitrary but fixed mapping respecting the topological ordering of
$\DG(\RS)$, i.e.~$\rk(\QQ_i)>\rk(\QQ_j)$ whenever $\QQ_j$ is reachable from
$\QQ_i$. We call $\rk(\C)$ the \emph{rank} of an SCC~$\C$.

\begin{defi}
\label{d:rank}
The \emph{rank of a dependency pair $s\rew t$}, denoted by $\rk(s\rew t)$, 
is the rank of $\PP$ such that ${s\rew t} \in \PP$. 
Let $u$ be a term and suppose there exists an SCC $\PP$ such that
$u^\sharp$ is \emph{not} a normal form with respect
to $\rsrew{\PP/\RS}$. The \emph{rank of the term $u$} is defined as follows:
\begin{equation*}
  \rk(u) \defsym \max\{\rk(s\rew t)\mid 
  \text{there exists $\sigma$ such that ${u^\sharp} \rssrew{\RS} {s\sigma}$} 
  \} \tpkt
\end{equation*}
\end{defi}

Observe that $\rk(u)$ need not be defined, although $u$ has a redex at the root position.
This is due to the fact that this redex need not be governed by a dependency
pair. On the other hand observe that if $u \not \in \NF(\PP/\RS)$ for
some SCC~$\PP$, then $\rk(u)$ is defined. Furthermore in this
case $\rk(u) > 0$ and $\dheight(u,\rsrew{\PP/\RS}) > 0$.

\begin{defi}
\label{d:sccheight}
We define the mapping $\sccheight$ from terms to
$\N \times \N$ as follows:
\begin{equation*}
  \sccheight(t)\defsym
  \begin{cases}
    (i, \dheight(t^\sharp,\rsrew{\PP_i/\RS})) &
    \text{if $\rk(t)$ is defined and $\rk(t)=i$} \tkom
    \\
    (0, 1) &
    \text{if $t^\sharp \in \NF(\PP/\RS)$ for all SCCs $\PP$
      and $\rt(t)$ is defined} \tkom
    \\
    (0, 0) &
    \text{otherwise} \tpkt
  \end{cases}
\end{equation*}
\end{defi}

In the following we write $\PP_i$ for an SCC with rank $i$. Note that any
SCC in $\DG(\RS)$ is uniquely defined by its rank.
We give some intuition for Definition~\ref{d:sccheight} that is made
precise in Lemma~\ref{lem:rank} below.
Consider a term $t$ and suppose $t$ is not in normal form with respect
to $\rsrew{\PP_i/\RS}$. Then the second component of $\sccheight(t)$
estimates the remaining rewrite steps with respect to $\PP_i$ modulo $\RS$.
The other cases in Definition~\ref{d:sccheight} take care of the possibility that
$t \in \NF(\QQ/\RS)$ for all SCCs $\QQ$ in $\DG(\RS)$.
Note that, if $\sccheight(t)=(i,j)$ with $i>0$, then $j>0$, as well.

\begin{exa}[continued from Example~\ref{ex:ancestors}]
\label{ex:sccheight}
There are two dependency pairs with respect to $\RSb$:
\begin{equation*}
\mm^\sharp(x)\rew\mpp^\sharp(\ma,x) \qquad
\mm^\sharp(x)\rew\ma^\sharp
\end{equation*}

The dependency graph of $\RSb$ contains no edges, so it only consists of two
(trivial) SCCs. Let $\rk(\mm^\sharp(x)\rew\mpp^\sharp(\ma,x))=1$, and
$\rk(\mm^\sharp(x)\rew\ma^\sharp)=2$, so
$\PP_1=\{\mm^\sharp(x)\rew\mpp^\sharp(\ma,x)\}$ and $\PP_2=\{\mm^\sharp(x)\rew\ma^\sharp\}$.

We now give the values of $\sccheight$ for all subterms of $t_1$ and
$t_2$. All function symbols in~$t_1$ and $t_2$ are defined, however the terms
$\ma^\sharp$ and $\mpp^\sharp(\ma,\ma)$ are normal forms with respect to
$\rsrew{\DP(\RSb)/\RSb}$. Therefore, $\sccheight(\ma)=\sccheight(\mpp(\ma,\ma))
=(0,1)$. On the other hand, we have 
\begin{align*}
\dheight(\mm^\sharp(\mm(\ma)),\rsrew{\PP_2/\RSb}) & = \dheight(\mm^\sharp(\ma),\rsrew{\PP_2/\RSb})
\\ & =\dheight(\mm^\sharp(\mpp(\ma,\ma)), \rsrew{\PP_2/\RSb})=1 \tpkt
\end{align*}
Thus, $\sccheight(\mm(\mm(\ma)))=\sccheight(\mm(\ma))=
\sccheight(\mm(\mpp(\ma,\ma)))=(2,1)$.
\end{exa}

We write $\glex$ for the lexicographic extension of the standard
order $>$ on the natural numbers.
\begin{lem}
\label{lem:rank}
Let $A \colon s \rsrew{\RS} t$, let $p\in\Pos(s)$, and let $q\in\Pos(t)$. 
Suppose that $q \in \child{p}{A}$.
Then $\sccheight(\atpos{s}{p})\gelex\sccheight(\atpos{t}{q})$. 
Let $p'$ be the redex position of $A$.
If $p=p'$, then $\sccheight(\atpos{s}{p}) \glex \sccheight(\atpos{t}{q})$. 
\end{lem}
\proof
By assumption $q \in \child{p}{A}$. Suppose that $\rt(\atpos{t}{q})$ is defined.
Otherwise let $\rt(\atpos{t}{q}) \in \CS$. Then 
by Definition~\ref{d:sccheight}, $\sccheight(\atpos{t}{q}) = (0,0)$
and the lemma is trivial.
Hence we assume $\rt(\atpos{t}{q})$ is
defined. Then by Lemma~\ref{lem:ancestortodp}
$(\atpos{s}{p})^\sharp  \rsgrew{\DP(\RS) \cup \RS} (\atpos{t}{q})^\sharp$.

Suppose further $p = p'$.
Then $(\atpos{s}{p})^\sharp \rsrew{\DP(\RS)} (\atpos{t}{q})^\sharp$
follows from the proof of Lemma~\ref{lem:ancestortodp}.
Thus $\rk(\atpos{s}{p})$ is defined and $\rk(\atpos{s}{p}) = i > 0$.
By Definition~\ref{d:rank} we have
$(\atpos{s}{p})^\sharp  \rsrew{\PP_j} (\atpos{t}{q})^\sharp$ for some
$j \leqslant i$. Thus $i > \rk(\atpos{t}{q})$, or $i=\rk(\atpos{t}{q})$ and 
\begin{equation*}
  \dheight((\atpos{s}{p})^\sharp,\rsrew{\PP_i/\RS}) >
\dheight((\atpos{t}{q})^\sharp,\rsrew{\PP_i/\RS}) \tpkt
\end{equation*}
The lemma follows.

Suppose otherwise $p \not= p'$. Then
\begin{equation*}
  \sccheight(\atpos{s}{p})\gelex\sccheight(\atpos{t}{q}) \tkom
\end{equation*}
by Definition~\ref{d:sccheight}. This concludes the proof of the lemma.
\qed

The simulating TRS $\RSsim$ is based on a mapping $\tr$
(see Definition~\ref{def:rstorspg} below)
such that $s\rsrew{\RS}t$ implies $\tr(s) \rstrew{\RSsim} \tr(t)$.
Essentially, $\tr(t)$ encodes $\sccheight(t')$ for all subterms $t'$
of $t$. This is done by transforming $t$ into a
term with an $a+1$-ary root symbol $\mg_i$, where $i$ is the first component
of $\sccheight(t)$, the first argument of $\mg_i$ encodes the second
component of $\sccheight(t)$, and the remaining arguments contain the
transformations of the direct subterms of $t$.

The main tool for achieving the simulation of a rewrite step $s\rsrew{\RS} t$
are rules which
create the progenies of the redex position $p'$ of the step. The set 
$\child{p'}{(s \rsrew{\RS} t)}$ contains at most 
$a^C$ many elements. We indicate how this behaviour is overapproximated
in derivations over $\RSsim$. For instance, if $a=2$ and $C=2$, we make
use of the following derivation:
\begin{equation*}
\mg_i(\ms(x),x_1,x_2) \rss
\mg_i(\mg_i(x,\mg_i(x,x_1,x_2),\mg_i(x,x_1,x_2)),\mg_i(x,\mg_i(x,x_1,x_2),\mg_i(x,x_1,x_2))) \tpkt
\end{equation*}

Recall the above definition of the TRS $\RS'$ computing the function 
$f$ defined in~\eqref{eq:scccomplexity}.
The definition of the simulating TRS $\RSsim$ employs
the TRS $\RS'$.

\begin{defi}
\label{d:1}
Consider the following TRS $\RSsim$, where
$0 \leqslant i \leqslant k$, $1 \leqslant i' \leqslant k$, and
$1 \leqslant j \leqslant a$.
\begin{alignat*}{2}
1_i\colon&\;& \mg_i(\ms(x),x_1,\ldots,x_a) &\rew \mtree_{i}(\ms^C(\Null),x,x_1,\ldots,x_a)
\\
2_{i'}\colon&\;& \mg_{i'}(x,x_1,\ldots,x_a) & \rew \mg_{i'-1}(\mf(\msize(\mg_0(\Null,x_1,\ldots,x_a))),x_1,\ldots,x_a)
\\
3_{i,j}\colon&\;& \msize(\mg_i(x,x_1,\ldots,x_a)) & \rew \md_a(\msize(x_j))
\\
4\colon&\;& \msize(\mc) &\rew \ms(\Null)
\\
5\colon&\;& \md_a(\ms(x)) &\rew \ms^a(\md_a(x))
\\
6\colon&\;& \md_a(\Null) &\rew \Null
\\
7\colon&\;& \mg_0(x,x_1,\ldots,x_a) &\rew \mc
\\
8_{i,j}\colon&\;& \mg_i(x,x_1,\ldots,x_a) &\rew x_j
\\
9\colon&\;& \mg(x) &\rew \mg_k(\mf(\msize(\mg_0(\Null,x,\ldots,x))),x,\ldots,x)
\\
10\colon&\;& \mz &\rew \mg_k(\mf(\msize(\mg_0(\Null,\mc,\ldots,\mc))),\mc,\ldots,\mc)
\\
11_{i}\colon&\;& \mtree_{i}(\Null,x,x_1,\ldots,x_a) &\rew \mg_i(x,x_1,\ldots,x_a)
\\
12_{i}\colon&\;& \mtree_{i}(\ms(y),x,x_1,\ldots,x_a) &\rew \mg_i(x,\mtree_{i}(y,x,x_1,\ldots,x_a),\ldots,\mtree_{i}(y,x,x_1,\ldots,x_a))
\tpkt
\end{alignat*}
These rules are augmented by $\RS'$ defining the function symbol $\mf$. 
Without loss of generality we can assume that the
signatures of $\RS'$ and $\RSsim$ are disjoint with the exception
of $\mf$ and the constructors $\ms$ and $\Null$.
\end{defi}

Observe that $\RSsim$ depends only on the constants $a$, $C$, $k$, and the function $f$.
Some comments: The rules $1_i$ ($0 \leqslant i \leqslant k$) are the main rules for the simulation
of the effects of a single step $s\rsrew{\RS}t$ in $\RSsim$. These rules have
already been motivated above.
The rules $2_{i'}$ ($1 \leqslant i' \leqslant k$)
simulate that each of the new positions $q$ created by $s\rsrew{\RS} t$
might be of rank $j$ ($i'>j$). Observe that by definition of the function $f$ we have
\begin{equation}
\label{eq:dg}
  f(\size{\atpos{t}{q}}) \geqslant \dheight((\atpos{t}{q})^\sharp,\rsrew{\C_j/\RS}) \tkom
\end{equation}
which explains the occurrence of the first argument of the right-hand side
of these rules.
The rules $3_{i,j}$--$6$ ($0 \leqslant i \leqslant k$, $1 \leqslant j \leqslant a$) 
define the function symbol $\msize$, that is, $\msize(s)$ reduces to
a numeral $\ms^l(\Null)$ such that $l \geqslant \size{s}$, see 
Lemma~\ref{lem:rspgsimulationhelp}(\ref{en:rspgsimulationhelp:3}) below.
The rules $7$--$8_{i,j}$ ($0 \leqslant i \leqslant k$, $1 \leqslant j \leqslant a$) 
make sure that any superfluous positions and copies of subterms created
by the rules of type $1_i$ can be deleted.
The rules $9$ and $10$ guarantee that the simulating derivation can be
started with a suitably small initial term. Note that it is in general not the
case that we have $\size{\tr(s)}\leqslant \size{s}$, compare
Definition~\ref{def:rstorspg} below.
Finally, the rules $11_{i}$--$12_{i}$ define the function symbols
$\mtree_{i}$ introduced by the rules $1_i$, which essentially unfold
a full $a$-ary tree using the function symbol~$\mg_i$.

Through a sequence of lemmata we show that the TRS $\RSsim$ indeed simulates
$\RS$ as requested. Let $\FS$, $\FSsim$ denote the signatures of
$\RS$ and $\RSsim$, respectively.

\begin{defi}
\label{def:rstorspg}
The mapping $\tr \colon \TA(\FS) \to \TA(\FSsim)$ is defined as follows.
Suppose $t=f(t_1,\ldots,t_n)$ and $\sccheight(t)=(i,l)$. Then we define:
\begin{equation*}
\tr(t) \defsym \mg_i(\ms^l(\Null),\tr(t_1),\ldots,\tr(t_n),\mc,\ldots,\mc)
\tpkt
\end{equation*}
\end{defi}

Note that for any constant $t$, we have $\tr(t) = \mg_i(\ms^l(\Null),\mc,\ldots,\mc)$ for some $i,l\in\N$. 
To simplify the presentation, we often compress sequences of $\mc$ as follows:
$\tr(t)=\mg_i(\ms^l(\Null),\omc)$. 
We exemplify the role played by the simulating TRS $\RSsim$ below.
\begin{exa}[continued from Example~\ref{ex:sccheight}]
Using the results of $\sccheight$ for all subterms of $t_1$ and $t_2$, we get
\begin{align*}
  \tr(\mm(\mm(\ma)))& =\mg_2(\ms(\Null),\mg_2(\ms(\Null),\mg_0(\ms(\Null),\omc),\mc),\mc)
\\
  \tr(\mm(\mpp(\ma,\ma)))&=\mg_2(\ms(\Null),\mg_0(\ms(\Null),\mg_0(\ms(\Null),\omc),\mg_0(\ms(\Null),\omc)),\mc)
\tpkt
\end{align*}
For this example, we use $\RSsim$ with the parameters $k=a=C=2$, and $f(n)=1$.
Hence, a suitable TRS $\RS'$ for defining $\mf$ consists of the single rewrite rule:
\begin{equation*}
  \mf(x)\rew\ms(\Null) \tpkt
\end{equation*}
The following derivation over $\RSsim\cup\RS'$ rewrites $\tr(t_1)$ into $\tr(t_2)$,
The underlined part in each term is the redex used in the next step in the
derivation. In each rewrite step using a rule from $\RSsim$, the applied rule
is indicated.
\begin{align*}
&\mg_2(\ms(\Null),\underline{\mg_2(\ms(\Null),\mg_0(\ms(\Null),\omc),\mc)},\mc)\\
&\qquad\rsrew{1_2}\mg_2(\ms(\Null),\underline{\mtree_{2}(\ms(\ms(\Null)),\Null,\mg_0(\ms(\Null),\omc),\mc)},\mc)\\
&\qquad\rsrew{12_{2}}\mg_2(\ms(\Null),\underline{\mg_2(\Null,\mtree_{2}(\ms(\Null),\Null,\mg_0(\ms(\Null),\omc),\mc),\mtree_{2}(\ms(\Null),\Null,\mg_0(\ms(\Null),\omc),\mc))},\mc)\\
&\qquad\rsrew{8_{2,1}}\mg_2(\ms(\Null),\underline{\mtree_{2}(\ms(\Null),\Null,\mg_0(\ms(\Null),\omc),\mc)},\mc)\\
&\qquad\rsrew{12_{2}}\mg_2(\ms(\Null),\mg_2(\Null,\underline{\mtree_{2}(\Null,\Null,\mg_0(\ms(\Null),\omc),\mc)},\mtree_{2}(\Null,\Null,\mg_0(\ms(\Null),\omc),\mc)),\mc)\\
&\qquad\rsrew{11_2}\mg_2(\ms(\Null),\mg_2(\Null,\underline{\mg_2(\Null,\mg_0(\ms(\Null),\omc),\mc)},\mtree_{2}(\Null,\Null,\mg_0(\ms(\Null),\omc),\mc)),\mc)\\
&\qquad\rsrew{8_{2,1}}\mg_2(\ms(\Null),\mg_2(\Null,\mg_0(\ms(\Null),\omc),\underline{\mtree_{2}(\Null,\Null,\mg_0(\ms(\Null),\omc),\mc)}),\mc)\\
&\qquad\rsrew{11_2}\mg_2(\ms(\Null),\mg_2(\Null,\mg_0(\ms(\Null),\omc),\underline{\mg_2(\Null,\mg_0(\ms(\Null),\omc),\mc)}),\mc)\\
&\qquad\rsrew{8_{2,1}}\mg_2(\ms(\Null),\underline{\mg_2(\Null,\mg_0(\ms(\Null),\omc),\mg_0(\ms(\Null),\omc))},\mc)\\
&\qquad\rsrew{2_2}\mg_2(\ms(\Null),\underline{\mg_1(\mf(\msize(\mg_0(\Null,\mg_0(\ms(\Null),\omc),\mg_0(\ms(\Null),\omc)))),\mg_0(\ms(\Null),\omc),\mg_0(\ms(\Null),\omc))},\mc)\\
&\qquad\rsrew{2_1}\mg_2(\ms(\Null),\mg_0(\underline{\mf(\msize(\mg_0(\Null,\mg_0(\ms(\Null),\omc),\mg_0(\ms(\Null),\omc))))},\mg_0(\ms(\Null),\omc),\mg_0(\ms(\Null),\omc)),\mc)\\
&\qquad\rsrew{\RS'}\mg_2(\ms(\Null),\mg_0(\ms(\Null),\mg_0(\ms(\Null),\omc),\mg_0(\ms(\Null),\omc)),\mc)
\tpkt
\end{align*}
\end{exa}\medskip

\noindent In the remainder of this section
we show that the derivational complexity of $\RSsim$, and thus of $\RS$, is
primitive recursive in $f$.
We define the equivalence $s \approx t$ on $\TA(\FSsim)$.
If $s = \mc$, then $t = \mc$. Otherwise if
$s = \mg_i(\ms^m(0),s_1,\ldots,s_a)$, then $t = \mg_{i'}(\ms^n(0),t_1,\dots,t_a)$,
such that $m,n \in \N$, $1 \leqslant i,i' \leqslant k$ and 
$s_j \approx t_j$ for all $1 \leqslant j \leqslant a$.

\begin{lem}
\label{lem:rspgsimulationhelp}
Let $i \in \{0,\dots,k\}$. Then the following properties of $\RSsim$ hold:
\begin{enumerate}[\em(1)]
\item \label{en:rspgsimulationhelp:1}
$\mg_i(\ms(x),x_1,\ldots,x_a)\rstrew{\RSsim}\mg_i(x,x_1,\ldots,x_a)$.
\item \label{en:rspgsimulationhelp:2}
$\mg_i(x,x_1,\ldots,x_a)\rstrew{\RSsim}\mc$
\item \label{en:rspgsimulationhelp:3}
For all ground terms $s$ such that $t \approx \tr(s)$, we have 
$\msize(t)\rstrew{\RSsim}\ms^{l}(\Null)$ where $l\geqslant\size{s}$.
\item \label{en:rspgsimulationhelp:4}
If $s\rstrew{\RS}t$ and $\tr(s)\rstrew{\RSsim}\tr(t)$ then for any
$n$-ary function symbol $f$, we have that
$\tr(f(u_1,\ldots,s,\ldots,u_n))\rstrew{\RSsim}\tr(f(u_1,\ldots,t,\ldots,u_n))$.
\end{enumerate}
\end{lem}
\proof
The first two assertions are obvious. 
We show the third part by induction on $\size{s}$. As $s$ is a ground term, 
$s=f(s_1,\ldots,s_n)$. Without loss of
generality we set $t=\mg_{0}(0,t_1,\ldots,t_n,\omc)$, 
where $s_j \approx \tr(t_j)$ for all $1 \leqslant j \leqslant n$.
If $\size{s}=1$, then $n=0$. Thus $\msize(t)\rew\msize(\mc)\rew\ms(\Null)$ 
by applying rules $8_{0,1}$ and $4$. Otherwise, suppose $\size{s}>1$. 
Then let $j$ be such that $\size{s_j}$ is maximal. By induction
hypothesis, we have $\msize(t_j)\rst \ms^{l_j}(\Null)$ with $l_j\geqslant\size{s_j}$. 
Hence, by applying rules $3_{0,j}$, $5$, and $6$, we obtain
\begin{equation*}
 \msize(t)\rew \md_a(\msize(t_j))\rss 
 \md_a(\ms^{l_j}(\Null)) \rss 
 \ms^{a\cdot l_j}(\Null) \tpkt 
\end{equation*}
Due to $a\cdot l_j\geqslant\size{s}$, property~(\ref{en:rspgsimulationhelp:3}) follows.

Finally, we show property~(\ref{en:rspgsimulationhelp:4}).  We set
$u \defsym f(u_1,\ldots,t,\ldots,u_n)$ and we suppose that
$\sccheight(f(u_1,\ldots,s,\ldots,u_n))=(i,m)$ and 
$\sccheight(f(u_1,\ldots,t,\ldots,u_n))=(i',m')$. Then we set
\begin{align*}
  s' & \defsym \tr(f(u_1,\ldots,s,\ldots,u_n))=\mg_i(m,\tr(u_1),\ldots,\tr(s),\ldots,\tr(u_n)) \\
  t' & \defsym \tr(f(u_1,\ldots,t,\ldots,u_n))=\mg_{i'}(m',\tr(u_1),\ldots,\tr(t),\ldots,\tr(u_n)) 
\tpkt
\end{align*}
By assumption, $f(u_1,\ldots,s,\ldots,u_n)$ rewrites to $f(u_1,\ldots,t,\ldots,u_n)$
by some derivation $A$. Moreover $\epsilon \in \child{\epsilon}{A}$ and thus by Lemma~\ref{lem:rank}
$(i,m)\gelex(i',m')$. 

If $i=i'$, then $m\geqslant m'$, and we have
\begin{align*}
s' & \rst \mg_i(m,\tr(u_1),\ldots,\tr(t),\ldots,\tr(u_n))
\\
  & \rss \mg_i(m',\tr(u_1),\ldots,\tr(t),\ldots,\tr(u_n)) = t'
\tpkt
\end{align*}
Here we apply the assumption $\tr(s) \rst \tr(t)$ in the first line and
property~(\ref{en:rspgsimulationhelp:1}) in the second.
Otherwise, $i>i'$ and we obtain the following derivation:
\begin{align*}
  s' & \rst \mg_i(m,\tr(u_1),\ldots,\tr(t),\ldots,\tr(u_n))
\\
  & \rss \mg_{i'}(\mf(\msize(t')),\tr(u_1),\ldots,\tr(t),\ldots,\tr(u_n))
\\
  & \rss \mg_{i'}(m,\tr(u_1),\ldots,\tr(t),\ldots,\tr(u_n)) = t'
\tpkt
\end{align*}
Here the second line follows by applying rules $2_i$ to $2_{i'+1}$ such that
$t' \approx \tr(u)$. In the third line, we 
firstly make use of property~(\ref{en:rspgsimulationhelp:3}) to conclude 
$\msize(t') \rst \ms^l(0)$ for some $l\geqslant\size{u}$. Secondly 
Definition~\ref{d:sccheight} yields $f(l) \geqslant m'$. Thus by~\eqref{eq:dg} 
we have $\mf(\ms^l(0)) \rss \ms^{f(l)}(0)$. Finally property~(\ref{en:rspgsimulationhelp:1})
is applied. 
This completes the proof of the lemma.
\qed

We arrive at the main lemma of this section.
\begin{lem}
\label{lem:rspgsimulation}
For any ground terms $s$ and $t$, $s\rsrew{\RS}t$ implies $\tr(s) \rstrew{\RSsim} \tr(t)$.
\end{lem}
\proof
Let $l\rew r$ be the rewrite rule applied in the step $s\rew t$. Then there exist some position
$p\in\FPos(s)$ and some substitution $\sigma$ such that $l\sigma=\atpos{s}{p}$ and 
$r\sigma=\atpos{t}{p}$. It is not difficult to see that 
there exists a position $q\in\FPos(\tr(s))$ such that $\tr(l\sigma)=\atpos{\tr(s)}{q}$ 
and $\tr(r\sigma)=\atpos{\tr(t)}{q}$.

First, we show $\tr(l\sigma)\rst \tr(r\sigma)$.
Let $\sccheight(l\sigma)=(i,m)$. Since $l$ is not a variable, we have $l=f(l_1,\ldots,l_n)$. Hence,
$\tr(l\sigma)=\mg_i(\ms^m(\Null),\tr(l_1\sigma),\ldots,\tr(l_n\sigma),\omc)$.
Since $\rt(l)$ is defined, we have $m>0$.
By rules $1_i$, $8_{i,1}$ and $12_{i}$, we have
\begin{equation*}
  \tr(l\sigma)\rst \mtree_{i}(\ms^{\depth{r}}(\Null),\ms^{m-1}(\Null),\tr(l_1\sigma),\ldots,\tr(l_n\sigma),\omc)
\tpkt
\end{equation*}

We show the following claim by induction on $\depth{u}$.
\begin{claim}
If $u\subterm r$, then 
$\mtree_{i}(\ms^{\depth{u}}(\Null),\ms^{m-1}(\Null), \tr(l_1\sigma),\ldots,\tr(l_n\sigma),\omc)
\rss \tr(u\sigma)$, where $\sccheight(l\sigma)=(i,m)$.
\end{claim}
Since $r\subterm r$ and $\depth{r}\leqslant C$, the claim entails
$\tr(l\sigma)\rst \tr(r\sigma)$. Applying
Lemma~\ref{lem:rspgsimulationhelp}(\ref{en:rspgsimulationhelp:4}) then
yields $\tr(s)\rst \tr(t)$ and the lemma follows. Hence, the remainder of
this proof is devoted to showing the claim.

In proof of the claim, it suffices to consider the interesting
case that $u\nprsubterm l$. 
Since $\Var(l) \supseteq \Var(r) \supseteq \Var(u)$,
$u$ is not a variable. Hence, $u = g(u_1,\ldots,u_{n'})$.
Let $\sccheight(u) = (i',m')$.
By induction hypothesis, for all $1 \leqslant j \leqslant n'$ we have
\begin{equation}
\label{eq:dg:2}
  \mtree_{i}(\ms^{\depth{u_j}}(\Null),\ms^{m-1}(\Null),
\tr(l_1\sigma),\ldots,\tr(l_n\sigma),\omc)\rss \tr(u_j\sigma) \tpkt
\end{equation}
Moreover, employing instances of the rules $8_{i,1}$ and $12_{i}$, we obtain:
\begin{align}
&  \mtree_{i}(\ms^{\depth{u}-1}(\Null),\ms^{m-1}(\Null),\tr(l_1\sigma),\ldots,\tr(l_n\sigma),\omc) \notag\\
&  \qquad\rss
   \mtree_{i}(\ms^{\depth{u_j}}(\Null),\ms^{m-1}(\Null),\tr(l_1\sigma),\ldots,\tr(l_n\sigma),\omc)
\label{eq:dg:3}
   \tpkt
\end{align}
From rule $12_i$ together with~\eqref{eq:dg:3} and~\eqref{eq:dg:2}, we obtain
\begin{equation*}
\mtree_{i}(\ms^{\depth{u}}(\Null),\ms^{m-1}(\Null), \tr(l_1\sigma),\ldots,\tr(l_n\sigma),\omc) \rss
\mg_i(\ms^{m-1}(\Null),\tr(u_1\sigma),\ldots,\tr(u_{n'}\sigma),\omc) \tkom
\end{equation*} 
employing Lemma~\ref{lem:rspgsimulationhelp}(\ref{en:rspgsimulationhelp:2}).
We distinguish two subcases for $i'$: either $i=i'$, or $i > i'$.  
(Note that $i'>i$ is impossible since $(i,m) \glex (i',m')$ due to Lemma~\ref{lem:rank}.)
Suppose $i'=i$, then $m > m'$ due to  Lemma~\ref{lem:rank}. We obtain
\begin{equation*}
\mg_i(\ms^{m-1}(\Null),\tr(u_1\sigma),\ldots,\tr(u_{n'}\sigma),\omc)\rss \tr(u\sigma) \tpkt
\end{equation*}
Here we use Lemma~\ref{lem:rspgsimulationhelp}(\ref{en:rspgsimulationhelp:1}).
Otherwise, if $i>i'$, 
from $\mg_i(\ms^{m-1}(\Null),\tr(u_1\sigma),\ldots,\tr(u_{n'}\sigma),\omc)$, 
we reach the term 
$\mg_{i'}(\mf(\msize(u')),\tr(u_1\sigma),\ldots,\tr(u_{n'}\sigma),
\omc)$ 
for a suitable $u'\approx \tr(u\sigma)$, applying rules $2_i$ to $2_{i'+1}$.
Thus by Lemma~\ref{lem:rspgsimulationhelp}(\ref{en:rspgsimulationhelp:1},\ref{en:rspgsimulationhelp:3}) 
and~\eqref{eq:dg}, we obtain the following derivation:
\begin{equation*}
  \mg_{i'}(\mf(\msize(u')),\tr(u_1\sigma),\ldots,\tr(u_{n'}\sigma),\omc) \rss
\mg_{i'}(\ms^{m'}(\Null),\tr(u_1\sigma),\ldots,\tr(u_{n'}\sigma),\omc) = \tr(u\sigma) \tpkt
\end{equation*}
This concludes the proof of the claim, and thus of the lemma.
\qed

Lemma~\ref{lem:rspgsimulation} yields that the length of any
derivation in $\RS$ can be estimated by the maximal derivation height with
respect to $\RSsim$. To extend this to measure the derivational complexity
function $\Dc{\RS}$ via the function $\Dc{\RSsim}$ we make use of the following
lemma; note that $\size{\mg^{\depth{t}}(\mz)}\leqslant\size{t}$.

\begin{lem}
\label{lem:rspgsimulationstart}
For any ground term $t$, we have $\mg^{\depth{t}}(\mz) \rssrew{\RSsim}\tr(t)$.
\end{lem}
\proof
We show the slightly more general assertion that if
$l \geqslant\depth{t}$, then $\mg^{l}(\mz) \rss \tr(t)$. We proceed
by induction on $l$.
Since $t$ is ground, $t=f(t_1,\ldots,t_n)$. Let $\sccheight(t)=(i,m)$.
We distinguish two cases: either
$l=0$ or $l>0$.
Assume $l=0$, then $\depth{t}=0$, hence $t$ is a constant.
We obtain the derivation
\begin{equation*}
  \mz\rew\mg_k(\mf(\msize(\mg_0(\Null,\omc))),\omc)
  \rss \mg_i(\mf(\msize(\mg_0(\Null,\omc))),\omc)
  \rss \mg_i(\mf(\ms^{l'}(\Null)),\omc)
  \rss \mg_i(\ms^{f(\size{t})}(\Null),\omc)
  \tkom
\end{equation*}
where $l'\geqslant\size{t}$.
Here we apply the rules $10$ and $2_k$ to $2_{i+1}$ in conjunction
with Lemma~\ref{lem:rspgsimulationhelp}(\ref{en:rspgsimulationhelp:3})
for the derivation $\mz \rss \mg_i(\mf(\ms^{l'}(\Null)),\omc)$.
Note that $\mg_0(\Null,\omc)\approx \tr(t)$ holds. Furthermore
$\mg_i(\mf(\ms^{l'}(\Null)),\omc) \rss \mg_i(\ms^{f(\size{t})}(\Null),\omc)$
is due to Lemma~\ref{lem:rspgsimulationhelp}(\ref{en:rspgsimulationhelp:1})
together with~\eqref{eq:dg}.

On the other hand assume $l >0$. It is easy to see that 
$\mg^{l}(\mz)\rss \mg^{\depth{t}}(\mz)$. Furthermore by an application of rule $9$,
we obtain
\begin{equation*}
  \mg^{\depth{t}}(\mz)\rew\mg_k(\mf(\msize(\mg_0(\Null,\mg^{\depth{t}-1}(\mz),\ldots,\mg^{\depth{t}-1}(\mz)))),\mg^{\depth{t}-1}(\mz),\ldots,\mg^{\depth{t}-1}(\mz))
  \tpkt
\end{equation*}
Note that $\mg(x)\rew\mg_k(\mf(\ldots),x,\ldots,x)\rss \mc$ by
Lemma~\ref{lem:rspgsimulationhelp}(\ref{en:rspgsimulationhelp:2}),
$\mz \rss \mc$, and for all $j$, $\mg^{\depth{t}-1}(\mz)\rss \tr(t_j)$ by
induction hypothesis. Thus the right-hand side of the above equation rewrites to
\begin{equation*}
 \mg_k(\mf(\msize(\mg_0(\Null,\tr(t_1),\ldots,\tr(t_n),\omc))),\tr(t_1),\ldots,\tr(t_n),\omc) \tkom 
\end{equation*}
which in turn rewrites to 
$\mg_i(\mf(\msize(s)),\tr(t_1),\ldots,\tr(t_n),\omc)$ 
for suitable $s$ with $s\approx \tr(t)$.
Finally, Lemma~\ref{lem:rspgsimulationhelp}(\ref{en:rspgsimulationhelp:1}),(\ref{en:rspgsimulationhelp:3}) 
and~\eqref{eq:dg} yield $\mg_i(\ms^m(\Null),\tr(t_1),\ldots,\tr(t_n),\omc)$.
This concludes the proof.
\qed

It remains to verify that $\RSsim$ is terminating and that $\Dc{\RSsim}$ is 
primitive recursive in $f$. This is non-trivial, due to the rules $1_i$.
\begin{thm}
\label{t:termination}
There exists a well-founded monotone algebra $\I$ such that $\I$ is
compatible with $\RSsim$ and for all $g \in \FSsim$, the function
$g_{\I}$ is primitive recursive in $f$. In particular $\RSsim$ is terminating
and $\Dc{\RSsim}$ is primitive recursive in $f$.
\end{thm}
\proof
The proof is given in the appendix.
\qed

We arrive at the main result of this section.
\begin{thm}
\label{thm:dgprecwall}
Let $\RS$ be a terminating TRS and let $f$ be the following function over $\N$:
\begin{equation*}
  f(n) \defsym \max (\{1\} \cup \{ \dheight(t^\sharp,\rsrew{\C/\RS})
  \mid \size{t} \leqslant n, \text{$\C$ is SCC of $\DG(\RS)$}\}) \tpkt
\end{equation*}
Then $\Dc{\RS}$ is primitive recursive in $f$. This upper bound is essentially
optimal if $f$ is at least linear.
\end{thm}
\proof
Let $t$ be a term. Without loss of generality we can assume that
$t$ is ground.
Due to Lemmata~\ref{lem:rspgsimulation} and \ref{lem:rspgsimulationstart}
we have the following inequalities. 
\begin{equation*}
  \dheight(t,\rsrew{\RS}) \leqslant \dheight(\tr(t),\rsrew{\RSsim})
  \leqslant \dheight(\mg^{\depth{t}}(\mz),\rsrew{\RSsim}) \tpkt
\end{equation*}
Note that $\size{\mg^{\depth{t}}(\mz)}\leqslant\size{t}$.
Hence for all $n$: $\Dc{\RS}(n)\leqslant\Dc{\RSsim}(n)$. 
Due to Theorem~\ref{t:termination}, $\Dc{\RSsim}$ is primitive recursive
in $f$. Thus $\Dc{\RS}$ is bounded by a function primitive recursive in
$f$.
It follows from Example~\ref{ex:dglowerbound}
that this bound is essentially optimal.
\qed

Consider any TRS $\RS$ whose termination can be shown by the basic dependency
pair method in conjunction with dependency graphs and some base technique
for each SCC of the dependency graph. Let $f$ be defined as in
Theorem~\ref{thm:dgprecwall}. As an example, if only polynomial
interpretations and MPO are used as base techniques, then both $f$ and
$\Dc{\RS}$ are bounded by primitive recursive functions.
Note that the derivational complexity induced by MPO (as a direct method) is primitive recursive~\cite{H92}.
By definition the primitive recursive functions are closed under
primitive recursion. Hence the complexity of the dependency pair method (in conjunction
with the dependency graph refinement) becomes negligible.

\section{Conclusion}
\label{sec:conclusion}

In this paper we have investigated the derivational complexity induced 
by the dependency pair method, where the object of our investigation is the 
standard formulation of the dependency pair method~\cite{AG00,HM05} together with natural refinements.

We have established the following results:
Firstly, for the basic dependency pair method (potentially using argument
filterings) the induced derivational complexity is triple 
exponentially bounded in the derivational complexity of the base technique 
used. For string rewrite systems we have an optimal exponential
upper bound and for the general case, we presented a double exponential
lower bound.
Secondly, if we consider the dependency pair method using
the usable rules refinement, then the induced derivational complexity
is primitive recursive in the derivational complexity of the base technique.
Here we have provided a nonelementary lower bound.
Finally, if we consider the dependency pair method in conjunction
with dependency graphs, then the induced derivational complexity is 
again primitive
recursive in the derivational complexity of the base techniques employed.
This result is essentially optimal.
It is worthy of note that this is the very first analysis of
the dependency pair method (without any dilutions) from a complexity analysis
point of view. 
It remains to clarify to what extent such results hold for other
notions of complexity.

As briefly mentioned in the introduction the \emph{derivational complexity} is
not the only measure of the complexity of a TRS suggested in the literature.
In particular, alternative approaches have been suggested by Choppy et al.~\cite{CKS:1989}, 
Cichon and Lescanne~\cite{CL:1992}, and Hirokawa and the first author~\cite{HM08a}. 
In~\cite{HM08a} the \emph{runtime complexity} with respect to a TRS is
defined as a refinement of the derivational complexity, by
restricting the set of admitted initial terms. This notion has
first been suggested in~\cite{CKS:1989}, 
where it is augmented by an \emph{average case}
analysis. Finally~\cite{CL:1992} studies the complexity of the 
functions \emph{computed} by a given TRS. 
This latter notion is often studied within \emph{implicit computational
complexity theory} (see~\cite{BMR:2009} for an overview).

We have chosen to present our results in terms of derivational complexity
as this simplifies the comparison to well-known results in this area. However,
it is easy to see that all upper bound results hold as well, if we
would study the runtime complexity of a TRS. 
Furthermore, the runtime complexity of a TRS is an invariant
cost model~\cite{LM09} and thus it is straightforward to 
rephrase our results in terms of the complexity of the function computed 
by the TRS in question. Let $f$ be a function
computable by a TRS $\RS$ and let $g$ denote a function
that grows at least linearly. Suppose the runtime complexity
of $\RS$ is bounded by $g(n)$. Then there exists a Turing machine
running in time polynomial in $g(n)$ that computes $f$~\cite{AM10}.
Thus our results also characterises the complexity of functions
computed by rewrite systems, whose termination has been shown by
the dependency pair method together with natural refinements.

From the original viewpoint of derivational complexity analysis, as an
analysis of the strength of termination methods, the implications of 
our results are easy to state.
For example, our results imply that the (technically simple) extensions 
of the dependency pair method with the dependency graph refinement
greatly increase the strength of the method. On the other hand our results
also provide limitations on the strength of the studied techniques.
For instance consider the following example.
\begin{exa}
Consider the TRS $\RStouzet$ introduced
by Touzet in~\cite{T:1998}.%
\footnote{This is example \textsf{Zantema\_04/z090} in the 
termination problems database, see \url{http://termcomp.uibk.ac.at/}.}
\begin{alignat*}{5}
\mb(\mu(x))&\rew\mb(\ms(x)) \hspace{1ex} & \ms(\mb(\ms(x)))&\rew\mb(\mt(x)) \hspace{2ex} 
& \mt(\mb(x))&\rew\mb(\ms(x)) \hspace{2ex} & \mt(\ms(x))&\rew\mt(\mt(x)) \\
\ms(\mb(x))&\rew\mb(\ms(\ms(\ms(x)))) & \ms(\mu(x))&\rew\ms(\ms(x)) 
& \mt(\mb(\ms(x)))&\rew\mu(\mt(\mb(x))) \hspace{2ex} & \mt(\mu(x))&\rew\mu(\mt(x))
\end{alignat*}
$\RStouzet$ encodes the Ackermann function~\cite{T:1998} and therefore
the derivational complexity function belongs to $\Ack(\Theta(n),0)$.
\end{exa}
Our results imply that any successful termination proof of $\RStouzet$
has to employ techniques that go beyond the basic dependency pair
method and the refinements studied here. Very recently, Sternagel and
Middeldorp presented in~\cite{SM:2008} 
an automatic termination proof of $\RStouzet$. Based on our work it is
indeed no surprise that this proof makes crucial use of 
an extension of the dependency pair method, the
dependency pair \emph{framework}~\cite{GTSF06,T07}.

Motivated by this and like-minded examples we have very recently started
investigations into the complexity induced by the 
dependency pair framework. A first result in this direction
shows that the complexity of the dependency pair framework may be
multiply recursive~\cite{MS:2011}. Furthermore, for a clearly defined
subset of processors, this bound is optimal.

In recent years (derivational) complexity results mainly focused on 
crafting new methods that induce low-complexity upper bounds, like for
example polynomial upper bounds. We exemplarily mention results by
Neurauter et al.\ studying the use of matrix interpretations to polynomially
bound the derivational complexity of TRSs~\cite{NZM:2010}. Moreover, in the
area of implicit computational complexity, Bonfante et al.\ study
the use of quasi-interpretations to characterise complexity classes like
$\Linspace$, $\Ptime$, or $\Pspace$~\cite{BMM:2011}. 
In the context of our results these classes are clearly of a 
low complexity.

With respect to this motivation our results
are arguably negative: our results clearly show that
the undiluted dependency pair method is not a suitable tool to yield
low complexity upper bound. Again it does not matter much 
whether we consider derivational complexity or runtime complexity: the
example given for the double exponential lower bound 
for the basic dependency pair method also shows a 
double exponential lower bound for the runtime complexity.

Recently a number of variants
of the dependency pair method have been proposed in the 
literature~\cite{AM09,HM08a,HM08b,MP08,MP09,NEG11,ZK10}. We believe that
our results can also be profitably employed in the crafting
of variants of the dependency pair method or framework 
in the context of \emph{polynomial} complexity analysis.
This will be subject to future work.

\section*{Acknowledgement}
We thank the reviewers for constructive suggestions that helped to improve the quality of the presentation
of the paper.

\clearpage
\appendix
\section{Termination of the Simulating TRS \texorpdfstring{$\RSsim$}{for Dependency Graphs}}

Recall the definition of the simulating TRS $\RSsim$ given in
Section~\ref{sec:dg}. In this appendix we define a well-founded monotone 
algebra $\I = (\N,>)$, where $>$ denotes
the usual order on the natural numbers. Termination of $\RSsim$ follows
as $\I$ is compatible with $\RSsim$. Furthermore, if the function
$f$, defined in~\eqref{eq:scccomplexity}, is primitive recursive, then 
$\I$ makes only use of primitive recursive interpretation functions.
The definition of $\I$ makes use of a family of fast growing functions,
defined below. This definition is parametrised in $d$. 
The exact value of the parameter $d$ will become clear from
the termination proof. To simplify the notation we assume the function $f$
is primitive recursive. Otherwise Definition~\ref{d:fast} has to be
replaced by a function hierarchy that is parametrised in $f$. 

\begin{defi}
\label{d:fast}
Let $d \geqslant 2$ be a given number. We define:
\begin{equation*}
  \Fast{0}(m) \defsym d^{m+1} \qquad
  \Fast{n+1}(m) \defsym \Fast{n}^{m+1}(m) \tpkt
\end{equation*}
\end{defi}

The following properties of the family of functions $\{ \Fast{n} \mid n \geqslant 0\}$
are easy to verify.
\begin{lem}
\label{l:fast}
Let $n$, $m$, $a$, and $b$ be natural numbers.
\begin{enumerate}[\em(1)]
\item \label{en:fast:1}
$\Fast{n}(a) \geqslant d^{a+1} \geqslant d \cdot a > a$.
\item \label{en:fast:2}
If $a > b$, then $\Fast{n}(a) > \Fast{n}(b)$.
\item \label{en:fast:3}
If $n > m$, then $\Fast{n}(a) > \Fast{m}(a)$ for
$a \geqslant 1$.
\item \label{en:fast:4}
$\Fast{m}(a+b) \geqslant \Fast{m}(a) + b$ and
$\Fast{m}(a+1) \geqslant 2\cdot \Fast{m}(a)$.
\item \label{en:fast:5}
Each function $\Fast{n}$ is primitive recursive.
\item \label{en:fast:6}
For every $n$-ary primitive recursive function $g$, there
exists a number $k$ such that for all numbers $m_1,\dots,m_n$:
$\Fast{k}(\max\{m_1,\dots,m_n\}) \geqslant g(m_1,\dots,m_n)$.
\qed
\end{enumerate}
\end{lem}

The next proposition is due to Hofbauer~\cite{H92b}.
\begin{prop} \label{p:algebra}
Let $\A=(\N,>)$ denote a weakly monotone algebra, compatible with a TRS $\RS$ and let
$p$ be a strictly monotone unary function on $\N$ such that for all $f \in \FS$
\begin{equation*}
  p(n) \geqslant f_{\A}(n,\dots,n) \qquad \text{for all $n \in \N$} \tpkt
\end{equation*}
Then we have $\Dc{\RS}(n) \leqslant p^n(0)$.
\qed
\end{prop}

Recall that $\FSsim$ denotes the signature of the TRS $\RSsim$.
By definition $\mf \in \FSsim$ and we assume that the function
$f$ is primitive recursive. The
rules $\RS'$ defining $\mf$ constitute a (terminating) subset of
$\RSsim$, c.f.~Definition~\ref{d:1}.
For the definition of the well-founded monotone algebra $\I$ it suffices
to define primitive recursive mappings $f_{\I}$ for all $f \in \FSsim$.
A complication is the definition of $\mf_\I$ as
the TRS $\RS'$ has only been defined implicitly above. However, following
the construction in~\cite{H92}, we conclude the existence of a
well-founded monotone algebra $\J$
compatible with $\RS'$ such that $\mf_{\J}$ is primitive recursive.
More precisely, without loss of generality we can assume that
there exists $\ell \in \N$ such that $\mf_{\J}(n) = \Fast{\ell}(n)$
and that $\ms_{\J}(n) = n + 1$ and $\Null_{\J} = 1$.

Preparing the definition of the well-founded monotone algebra $\I$, we define
the interpretation functions $\mf_{\I}$, $\ms_{\I}$, and $\Null_{\I}$ as follows:
\begin{equation}
\label{eq:I:g}
    \mf_{\I}(n) \defsym \Fast{\ell}(n) \qquad
    \ms_{\I}(n) = n + 1 \qquad
    \Null_{\I} = 1
    \tpkt
\end{equation}
The next definition gives the mappings associated to the function
symbols $\mg_i$ ($0 \leqslant i \leqslant k$).
Let $d \geqslant \max\{C+2,a+1\}$.
\begin{equation}
\label{eq:I:fi}
  (\mg_i)_{\I}(n,x_1,\dots,x_a) \defsym \Fast{\ell+2i}^{d^{n+1} \cdot (C+1)}(n+x_1+\cdots+x_a) \tpkt
\end{equation}

Before we continue the definition of $\I$ we give the following 
auxiliary result. Let $\alpha$ denote an arbitrary assignment. Let $x$ be a variable and let $\ox$ abbreviate $\eval{\alpha}{x}$.
\begin{lem}
\label{l:2i'}
Let $\alpha$ be an assignment such that for all $x \in \VS$,
$\alpha(x) \geqslant 1$. Then there exists $q\in\N$ such that
$\Fast{q}(\on + \ox_1 + \cdots + \ox_a) \geqslant \eval{\alpha}{\mg_i(n,x_1,\dots,x_a)}$.
\end{lem}
\proof
By definition $\eval{\alpha}{\mg_i(n,x_1,\dots,x_a)} = \Fast{\ell+2i}^{d^{\on+1} \cdot (C+1)}(\on+\ox_1+\cdots+\ox_a)$, we set $p \defsym \ell + 2i$ and abbreviate
$\on + \ox_1 + \cdots + \ox_a$ as $\on + \ox$. 
Due to Lemma~\ref{l:fast}(\ref{en:fast:1}) and the assumption
on $d$, we have $\Fast{1}(\on+\ox) \geqslant d^{\on+2+\ox} \geqslant d^{\on+2} + \ox
> d^{\on+1} \cdot (C+1) + \ox$
for $\on \geqslant 1$. In sum, we obtain:
\begin{align*}
   \Fast{p+2}(\on + \ox) & \geqslant \Fast{p+1} \circ \Fast{p+1}(\on + \ox)\\
   & \geqslant \Fast{p+1} \circ \Fast{1}(\on + \ox)\\
   & \geqslant \Fast{p+1} (d^{\on+1}\cdot (C+1) + \ox)\\ 
   & \geqslant \Fast{p}^{d^{\on+1}\cdot (C + 1) +1} (d^{\on+1}\cdot (C+1)+\ox) \\
   & > \Fast{p}^{d^{\on+1} \cdot (C+1)}(\on+\ox) \tpkt
\end{align*}
Hence the lemma follows, if we set $q \defsym p +2$.
\qed

The next definition gives the mappings associated to the function
symbols $\mtree_i$~($0 \leqslant i \leqslant k$).
\begin{equation}
\label{eq:I:Mi}
  (\mtree_i)_{\I}(m,n,x_1,\dots,x_a) \defsym \Fast{\ell+2i}^{d^{n+2} \cdot (m +1)}(x+x_1+\cdots+x_a)
  \tpkt
\end{equation}

The interpretation functions given in~\eqref{eq:I:g}--\eqref{eq:I:Mi} are sufficient
to prove the main result of this appendix.
\begin{thmcont}{t:termination}
There exists a well-founded monotone algebra $\I$, such that $\I$ is
compatible with $\RSsim$ and for all $g \in \FSsim$, the function 
$g_{\I}$ is primitive recursive in the parameter function $f$. 
In particular $\RSsim$ is terminating.
\end{thmcont}
\proof
Without loss of generality we can assume that $f$ is primitive recursive.
Otherwise a straightforward extension of Definition~\ref{d:fast} suffices to
prove the more general proposition.

Set $\I = (\N - \{0\},>)$ and recall that in~\eqref{eq:I:g}, \eqref{eq:I:fi}, 
and~\eqref{eq:I:Mi} the mappings $\mf_{\I}$, $\ms_{\I}$, $\Null_{\I}$, $(\mg_i)_{\I}$
and $(\mtree_i)_{\I}$ have been defined, where $0 \leqslant i \leqslant k$ and
$0 \leqslant j \leqslant C$ holds.
We extend these definitions, by setting $\mc_{\I} \defsym 3$ and $\msize_{\I}(n) \defsym n$.
Hence it remains to consider the mappings $\md_\I$, $\mg_{\I}$, and $\mz_{\I}$.
Based on Lemma~\ref{l:2i'} it is not difficult to define
suitable interpretations such that the 
rules $3_{i,j}$---$10$ are strictly decreasing with respect to $\I$. We leave
these definitions to the reader.

We write $f \circ g(n)$ for the function composition $f(g(n))$ and we 
abbreviate $\ox_1 + \cdots + \ox_a$ as $\overline{x}$. 
In proving compatibility, we restrict our attention to the (families of) rules $1_i$,
$11_i$, $12_i$ ($i \in \{0,\dots,k\}$) and $2_{i'}$ ($i' \in \{1,\dots,k\}$).
Let $i$ be arbitrary, but fixed. 

Consider the rule $1_i$: 
\begin{equation*}
  \mg_i(\ms(n),x_1,\ldots,x_a) \rew \mtree_{i}(\ms^C(0),x,x_1,\ldots,x_a) 
  \tpkt
\end{equation*}
Due to Lemma~\ref{l:fast}(\ref{en:fast:1}) we obtain (for an arbitrary assignment $\alpha$):
\begin{align*}
  \eval{\alpha}{\mg_i(\ms(n),x_1,\ldots,x_a)} 
  & = \Fast{\ell+2i}^{d^{\on+2} \cdot (C+1)}(\on+1+\overline{x})\\
  & > \Fast{\ell+2i}^{d^{\on+2} \cdot (C+1)}(\on+\overline{x})\\
  & = \eval{\alpha}{\mtree_i(\ms^C(0),n,x_1,\dots,x_a)} \tpkt
\end{align*}

Consider the rule $11_i$: 
\begin{equation*}
  \mtree_{i}(0,x,x_1,\ldots,x_a) \rew \mg_i(x,x_1,\ldots,x_a) \tpkt
\end{equation*}
Then we obtain:
\begin{align*}
\eval{\alpha}{\mtree_i(0,n,x_1,\dots,x_a)} & = \Fast{p}^{d^{\on+2}}(\on+\ox_1+\cdots+\ox_a)
\\ 
& > \Fast{p}^{d^{\on+1} \cdot (C+1)}(\on+\ox_1+\cdots+\ox_a)
\\
& = \eval{\alpha}{\mg_i(n,x_1,\dots,x_a)} \tkom
\end{align*}
where we use Lemma~\ref{l:fast}(\ref{en:fast:1}) together
with the fact $d > C+1$. 

Consider the rule $12_i$:
\begin{equation*}
  \mtree_{i}(\ms(y),x,x_1,\ldots,x_a) \rew 
  \mg_i(x,\mtree_{i}(y,x,x_1,\ldots,x_a),\ldots,\mtree_{i}(y, x,x_1,\ldots,x_a))
  \tpkt
\end{equation*}
We obtain:
\begin{align*}
\eval{\alpha}{\mtree_i(\ms(m),n,x_1,\dots,x_a)} &=  \Fast{p}^{d^{\on+2} \cdot (\om+2)}(\on+\ox)
\\
& > \Fast{p}^{d^{\on+2} \cdot (\om+1) + d^{\on+2} - d^{\on+1} + 1}(\on + \ox) 
\\
& = \Fast{p}^{d^{\on+2} \cdot (\om+1) + d^{\on+1} \cdot (d-1) + 1}(\on + \ox) 
\\
& \geqslant \Fast{p}^{d^{\on+2} \cdot (\om+1) + d^{\on+1} \cdot (C+1) + 1}(\on + \ox)
\\
& = \Fast{p}^{d^{\on+1} \cdot (C+1)} \circ \Fast{p} \circ \Fast{p}^{d^{\on+2} \cdot (\om+1)} (\on + \ox) 
\tpkt
\end{align*}
Due to~\eqref{eq:I:Mi} $\Fast{p}^{d^{\on+2} \cdot (\om+1)} (\on + \ox) 
= \eval{\alpha}{\mtree_i(m,n,x_1,\dots,x_a)}$. Thus due to
Lemma~\ref{l:fast}(\ref{en:fast:1}) and $d \geqslant a+1$, we obtain:
$\Fast{p} \circ \Fast{p}^{d^{\on+2} \cdot (\om+1)} (\on + \ox) \geqslant 
\on + a \cdot \eval{\alpha}{\mtree_i(m,n,x_1,\dots,x_a)}$.
Moreover, due to~\eqref{eq:I:fi}: 
\begin{gather*}
  \Fast{p}^{d^{\on+1} \cdot (C+1)} (\on+a \cdot \eval{\alpha}{\mtree_i(m,n,x_1,\dots,x_a)}) =
\\
  \eval{\alpha}{\mg_i(n,\mtree_i(m,n,x_1,\dots,x_a),\dots,\mtree_i(m,n,x_1,\dots,x_a))} 
  \tpkt
\end{gather*}
In sum we obtain 
\begin{gather*}
\Fast{p}^{d^{\on+1} \cdot (C+1)} \circ \Fast{p} \circ \Fast{p}^{d^{\on+2} \cdot (\om+1)} (\on + \ox) \geqslant
\Fast{p}^{d^{\on+1} \cdot (C+1)} (\on+a \cdot \eval{\alpha}{\mtree_i(m,n,x_1,\dots,x_a)}) =
\\
= \eval{\alpha}{\mg_i(n,\mtree_i(m,n,x_1,\dots,x_a),\dots,\mtree_i(m,n,x_1,\dots,x_a))} 
  \tkom
\end{gather*}
where we employ Lemma~\ref{l:fast}(\ref{en:fast:2}).

Finally, consider the family of rules $(2_{i'})_{1 \leqslant i' \leqslant k}$ and 
let $i' \in \{1,\dots,k\}$ be arbitrary, but fixed. Consider the rule $2_{i'}$:
\begin{equation*}
\mg_{i'}(n,x_1,\ldots,x_a) \rew \mg_{i'-1}(\mg(\msize(\mg_{0}(\Null,x_1,\ldots,x_a))),x_1,\ldots,x_a)  
\tpkt
\end{equation*}
Set $p \defsym \ell+2i'$, hence $p-2 = \ell + 2(i'-1)$.
Recall that $d \geqslant C +2 \geqslant 4$.

We obtain for any assignment $\alpha$ such that $\alpha(x) \geqslant 1$ for all
$x \in \VS$:
\begin{align*}
\eval{\alpha}{\mg_{i'}(n,x_1,\ldots,x_a)}
& = \Fast{p}^{d^{\on+1} \cdot (C+1)}(\on+\overline{x}) \\
& > \Fast{p} \circ \Fast{p} \circ \Fast{p}^{d \cdot (C+1)}(\on+\overline{x}) \\
& > \Fast{p} \circ \Fast{p} \circ \Fast{p-2}^{d^{\on+2} \cdot (C+1)}(\on+\overline{x})\\
& \geqslant \Fast{p} \circ \Fast{\ell} \circ \Fast{p-2}^{d^{\on+2} \cdot (C+1)}(1+\overline{x})\\
& \geqslant \Fast{p-2}^{d^{(\Fast{\ell} \circ \Fast{p-2}^{d^2 \cdot (C+1)}(1+\overline{x}))+1} \cdot (C+1)} \circ \Fast{\ell} \circ \Fast{p-2}^{d^{\on+2} \cdot (C+1)}(1+\overline{x})\\
& \geqslant \Fast{p-2}^{d^{\Fast{\ell}(\Fast{\ell}^{d^2 \cdot (C+1)}(1+\overline{x}))+1} \cdot (C+1)}(\Fast{\ell}(\Fast{\ell}^{d^2 \cdot (C+1)}(1+\overline{x})) + \overline{x})\\
& = \eval{\alpha}{\mg_{i'-1}(\mf(\msize(\mg_{0}(\Null,x_1,\ldots,x_a))),x_1,\ldots,x_a)}
\tpkt
\end{align*}
In lines~3 and 5 we apply slight variants of the proof of
Lemma~\ref{l:2i'}, and in line~6 we apply Lemma~\ref{l:fast}(\ref{en:fast:4}).

This completes the proof of compatibility for the crucial families of rules $1_i$,
$11_i$, $12_i$ ($i \in \{0,\dots,k\}$) and $2_{i'}$ ($i' \in \{1,\dots,k\}$). Hence
the theorem follows.
\qed
\
\end{document}